\documentclass[prx, twocolumn, showpacs]{revtex4-1}
\usepackage{float}
\usepackage{bbm}
\usepackage{epsfig}
\usepackage{epstopdf}
\usepackage{graphicx}
\usepackage{amsmath,amssymb}
\usepackage{physics}
\usepackage{color}
\usepackage{hyperref}
\usepackage[caption=false]{subfig}
\usepackage{siunitx, booktabs}
\usepackage{diagbox, eqparbox, hhline}
\newcolumntype{P}[1]{>{\centering\arraybackslash}p{#1}}

\begin{document}

\title{Interplay of exchange and superexchange in triple quantum dots}

\author{Kuangyin Deng}
\author{Edwin Barnes}
\email{efbarnes@vt.edu}
\affiliation{Department of Physics, Virginia Tech, Blacksburg, VA 24061, USA}

\begin{abstract}
Recent experiments on semiconductor quantum dots have demonstrated the ability to utilize a large quantum dot to mediate superexchange interactions and generate entanglement between distant spins. This opens up a possible mechanism for selectively coupling pairs of remote spins in a larger network of quantum dots. Taking advantage of this opportunity requires a deeper understanding of how to control superexchange interactions in these systems. Here, we consider a triple-dot system arranged in linear and triangular geometries. We use configuration interaction calculations to investigate the interplay of superexchange and nearest-neighbor exchange interactions as the location, detuning, and electron number of the mediating dot are varied. We show that superexchange processes strongly enhance and increase the range of the net spin-spin exchange as the dots approach a linear configuration. Furthermore, we show that the strength of the exchange interaction depends sensitively on the number of electrons in the mediator. Our results can be used as a guide to assist further experimental efforts towards scaling up to larger, two-dimensional quantum dot arrays.
\end{abstract}


\maketitle

\section{Introduction}\label{sec:Intro}
Semiconductor quantum dot spin systems are promising platforms for quantum computation because of their small scale, fast controllability, and long coherence times ~\cite{Loss1998,Hanson2007,Zwanenburg2013}. Qubits based on electron spins in quantum dots come in several varieties, including ones based on individual electron spins ~\cite{Loss1998,Elzerman2004,Koppens2006}, two-electron singlet-triplet qubits ~\cite{Petta2005,Maune2012}, and three-electron resonant exchange qubits ~\cite{Laird2010,Medford2013a,Medford2013,Eng2015} and hybrid qubits ~\cite{Shi2012,Koh2012,Cao2016}. In the past few years, there has been rapid progress in improving gate fidelities and in scaling up to larger quantum dot spin arrays ~\cite{Veldhorst2014,Veldhorst2015a,Kim2014,Kim2015b,Takeda2016a,Nichol2017,Ito2018,Watson2018,Zajac2018,Mills2019,Sigillito2019,Huang2019,Kandel2019,Dehollain2019,Lawrie2020}. There have also been remarkable advances in creating long-distance spin-spin interactions using superconducting resonators \cite{Viennot2015,Samkharadze2018,Mi2018,Landig2018} or a large multi-electron quantum dot as a mediator of superexchange interactions \cite{Mehl2014,Srinivasa2015,Baart2017,Croot2018,Malinowski2019}. As for most approaches to quantum computing, one of the current challenges in this field is to determine suitable, scalable architectures that achieve high connectivity and controllability while maintaining long coherence times.

In all these types of quantum dot spin qubits, exchange interactions play a central role, either as a main driver of entanglement generation or as the primary single-qubit control mechanism in the case of qubits based on the spin states of two or three electrons. The exchange energy between two spins is defined as the energy splitting between the triplet state with $S_z=0$ and the singlet state. For two electrons in the same quantum dot, the exchange energy can be positive or negative ~\cite{Wagner1992,Baruffa2010a,Zumbuhl2004,Mehl2014a}. For example, it has been shown that the ground state oscillates between a singlet and a $S_z=0$ triplet as the magnetic field strength is tuned, even in the absence of spin-orbit coupling \cite{Wagner1992,Mehl2014a}. It is also possible to have a triplet ground state without a magnetic field by  increasing the number of electrons in the dot \cite{Deng2018}. On the other hand, triplet ground states for electrons inhabiting two different dots are rare due to the interplay between the Coulomb interaction and the tunneling strength. This means that the effective exchange energy between two electrons on different dots is normally positive, which in turn limits the types of control schemes that can be employed to perform logic gates or dynamical decoupling \cite{Wang2012,Kestner2013,Wang2014}. 

However, it has been shown in recent experiments that the behavior of the exchange energy can be very different if there is a big multielectron quantum dot in the system ~\cite{Martins2017,Malinowski2018,Malinowski2019}. These experiments reported negative exchange energies due to contributions from electrons in the higher orbitals of the big quantum dot. It has also been experimentally demonstrated that multielectron quantum dots can be used to mediate strong superexchange interactions between spins that do not interact directly \cite{Malinowski2019}. Together, these findings suggest that architectures based on arrays of smaller one-electron dots interspersed with larger multielectron dots may be a promising route to scaling up to larger quantum processors \cite{Malinowski2017thesis}. For instance, one could imagine a square 2d array of single-electron quantum dots with a large multielectron mediator at the center of each plaquette. A key outstanding question is whether one can selectively interact pairs of spins coupled to the same mediator by adjusting detunings and tunnel barriers.

\begin{figure*}[!tbp]
\centering
\subfloat[Case 1]
{{\includegraphics[trim=0cm 0cm 0cm 0cm, clip=true,width=4.5cm, angle=0]{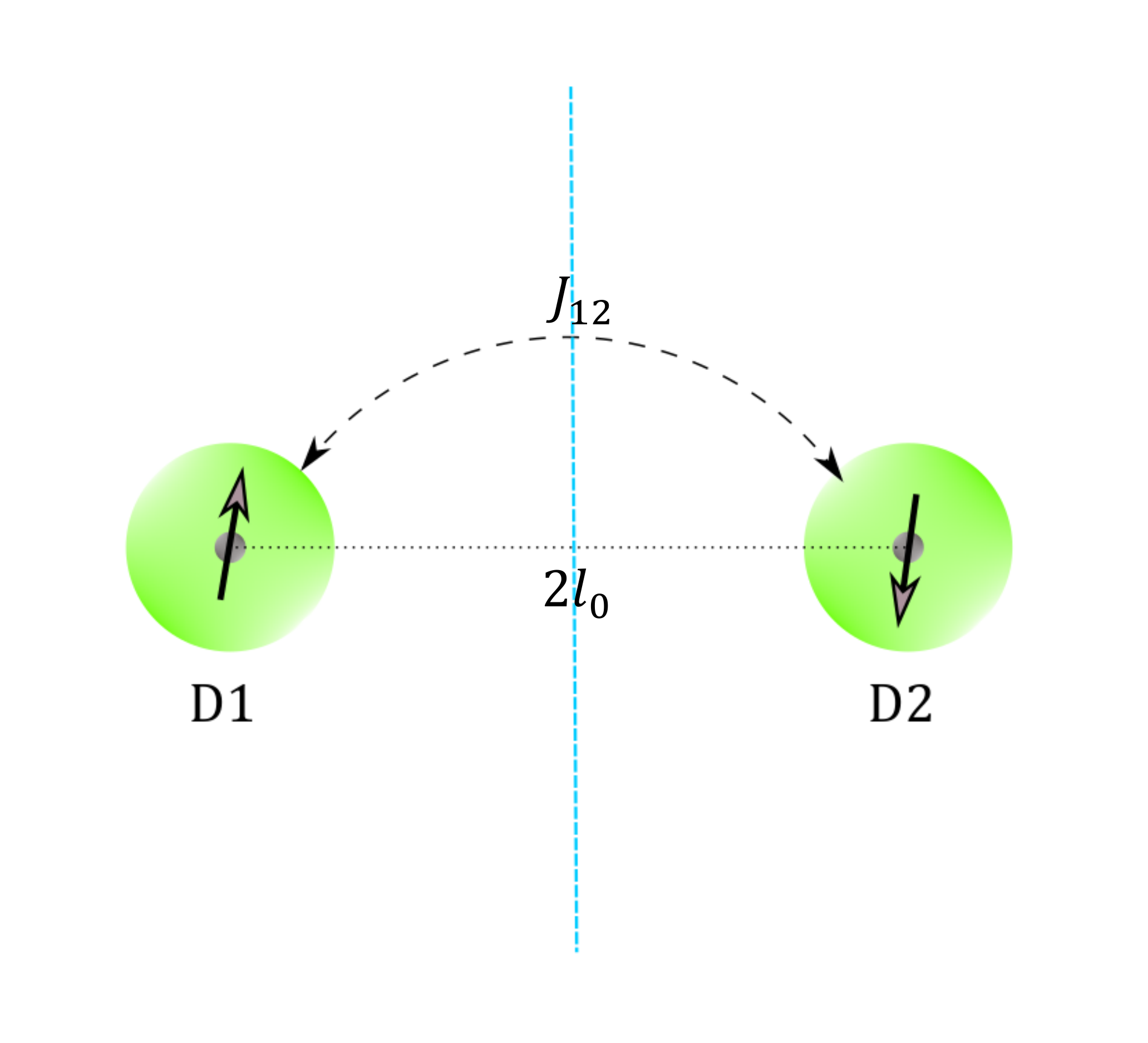}}}
\subfloat[Case 2]
{{\includegraphics[trim=0cm 0cm 0cm 0cm, clip=true,width=4.5cm, angle=0]{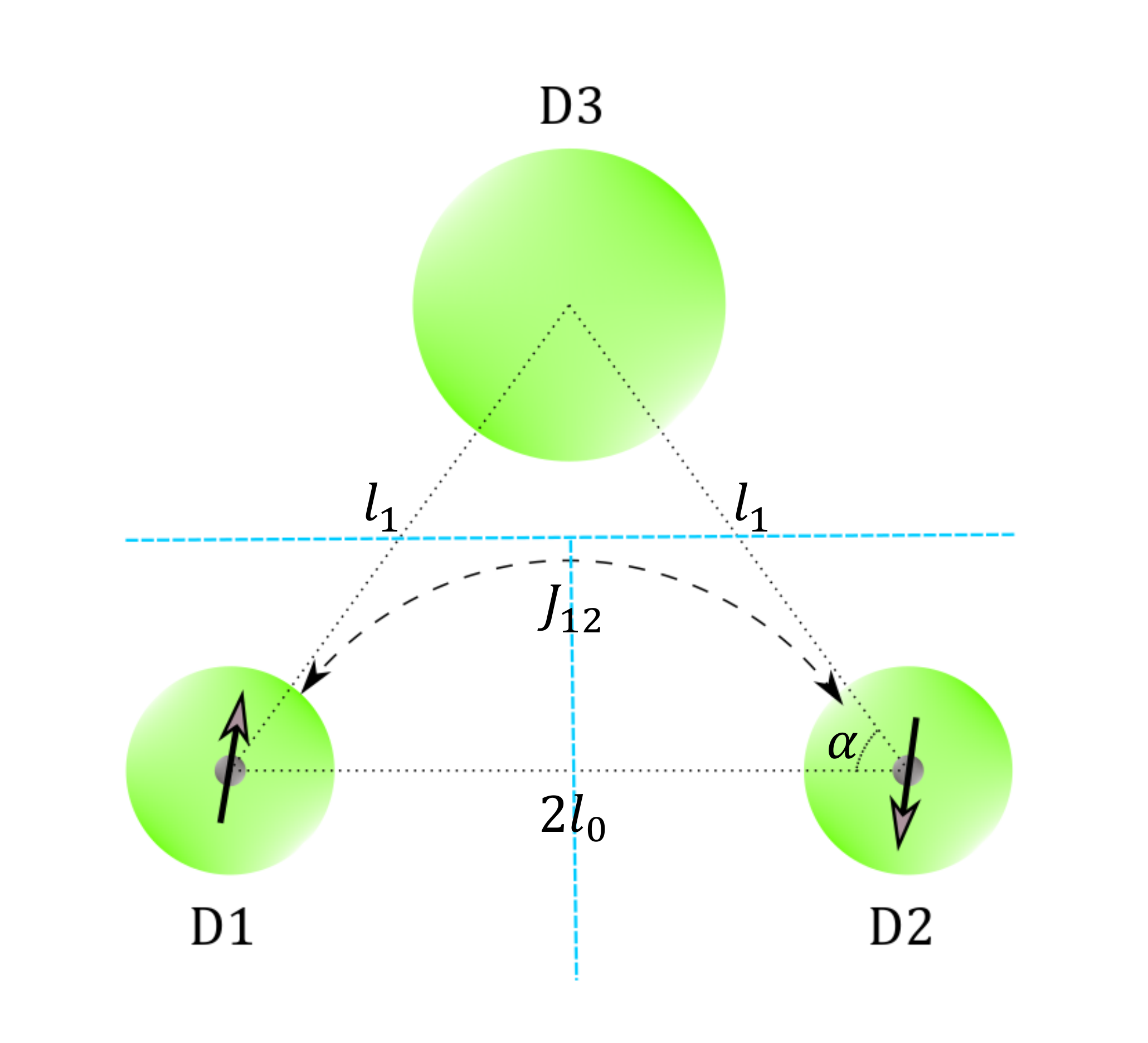}}}
\subfloat[Case 3]
{{\includegraphics[trim=0cm 0cm 0cm 0cm, clip=true,width=4.5cm, angle=0]{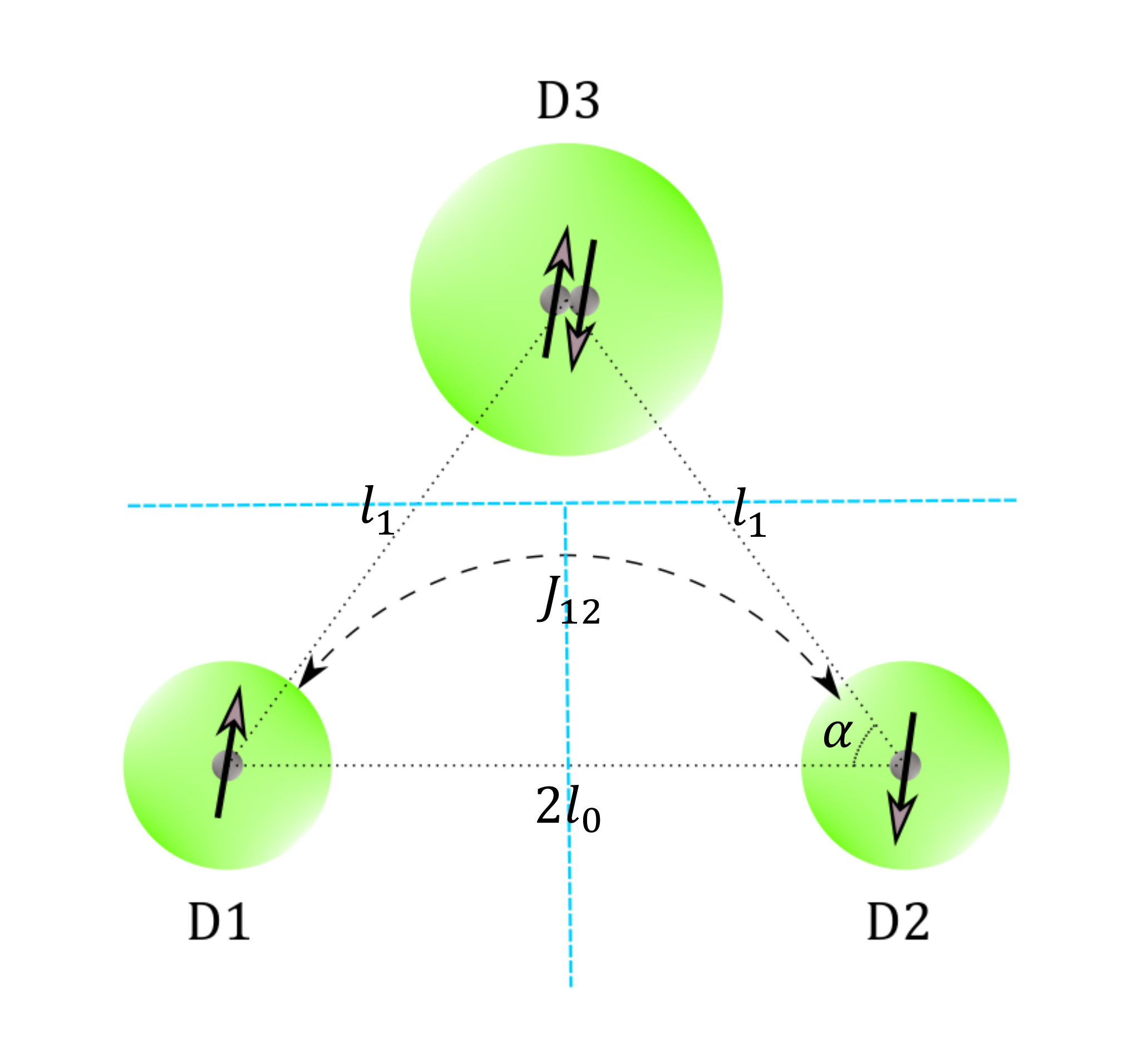}}}
\subfloat[Case 4]
{{\includegraphics[trim=0cm 0cm 0cm 0cm, clip=true,width=4.5cm, angle=0]{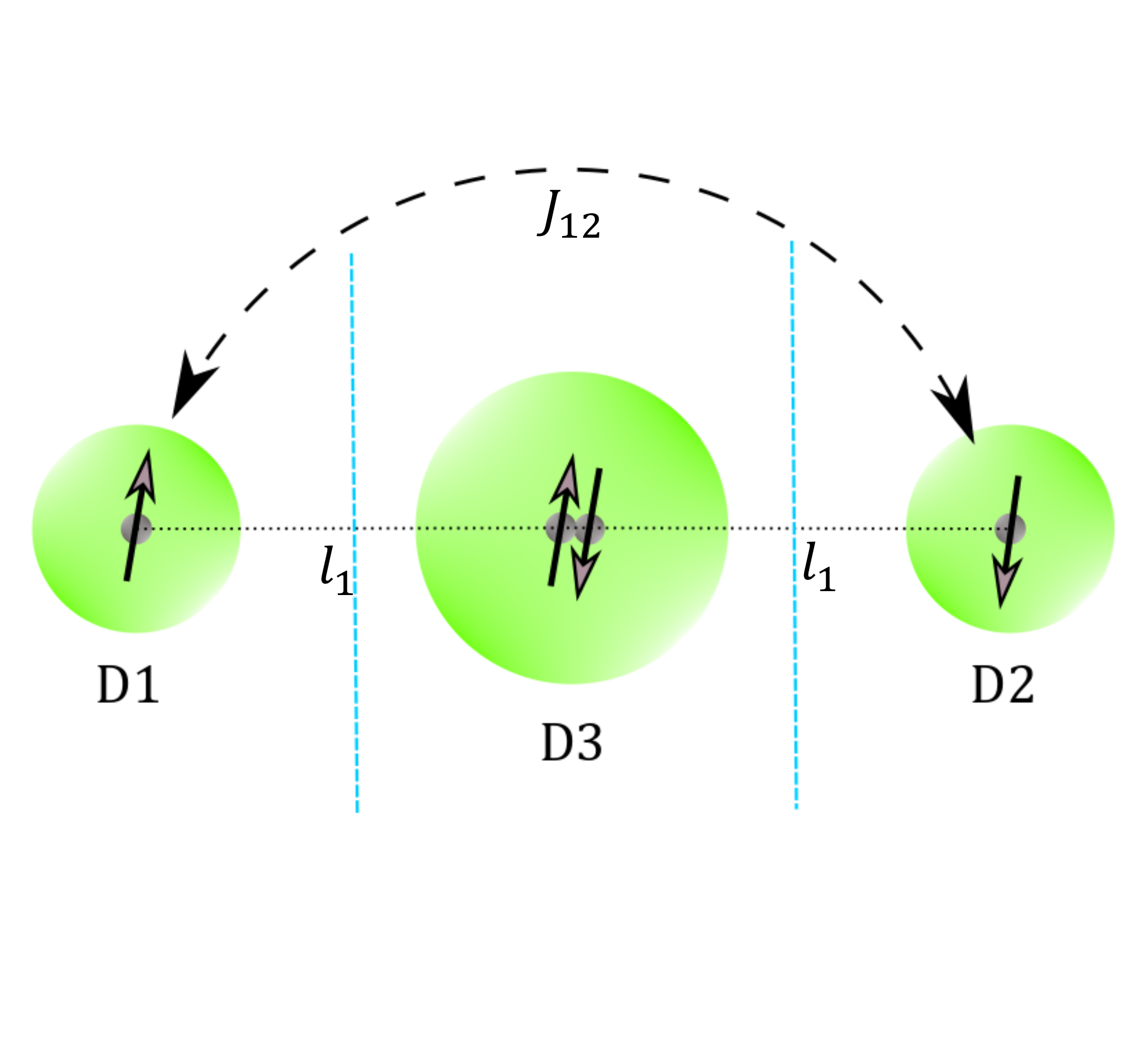}}}
\caption{Four different quantum dot configurations studied in this work. D1 and D2 are small dots about $12.5$ nm in radius, which corresponds to a confinement energy of $\hbar\omega_0=7.28$ meV. The $x$-axis is defined to pass through the centers of both D1 and D2. D3 is the mediator quantum dot, which is taken to have a radius of about $17.5$ nm, which corresponds to a confinement energy of $\hbar\omega_1=\frac{1}{2}\hbar\omega_0=3.64$ meV. The $y$-axis passes through the center of D3. Here, we fix $2l_0=56$ nm for cases 1, 2 and 3. We distinguish different triangular geometries in cases 2 and 3 by angle $\alpha$; the distance between the mediator and small dots, $l_1$, is a function of $\alpha$. The blue dashed lines in every case are the potential separation lines described in the main text. (a) The two-dot system. $J_{12}$ is the exchange coupling between two dots, which are separated by $2l_0$. (b) The triangular system with two electrons. We adjust the detuning $\Delta$ to keep D3 empty. (c) The triangular system with four electrons. We confine two electrons in the big dot by adjusting the detuning parameter $\Delta$. (d) Linear three-dot with four electrons. This can be thought of as a limit of case 3 in which we fix $l_1$ and rotate the two small dots D1 and D2 with respect to D3 until they are on a line. $l_1$ in this case is still a function of $\alpha$ as in case 3, and we also use $\alpha$ to adjust the inter-dot separation in this case.}\label{fig:1}
\end{figure*}

In this work, we take a first step toward addressing this question by investigating the interplay of normal exchange and superexchange in triple quantum dot systems where one of the dots is a large mediator. Using configuration interaction (CI) calculations, we explore how these two types of exchange evolve as the geometry and electron number in the dots are varied. Specifically, we compare four different cases: (i) a double quantum dot without the mediator, a triple quantum dot in a triangular configuration with (ii) two electrons or (iii) four electrons, and (iv) a triple quantum dot in a linear arrangement with the mediator in the middle. These cases are summarized in Fig.~\hyperref[fig:1]{\ref*{fig:1}}. In the triangular configuration cases, we find that the mediator gives rise to a modest superexchange interaction when it is not occupied, but when two electrons are added to the mediator, this interaction becomes orders of magnitude stronger. We also find that the effective exchange exhibits non-monotonic behavior as the mediator moves away from the two smaller dots. In the linear configuration case, we find that including the mediator leads to a still stronger exchange coupling, along with a substantial extension of the interaction distance of the two remote spin qubits.

This paper is organized as follows. In Sec.~\ref{sec:CI_two_dots}, we give details of the system Hamiltonian and the CI approach we use. We also compute the exchange energy for two electrons in two dots, which is used as a reference to compare against the triple-dot configurations. In Sec.~\ref{sec:Triangular_two_electrons}, we investigate superexchange in the triangular dot configuration with two electrons. In Sec.~\ref{sec:Triangular_four_electrons}, we study the effect of adding two electrons to the mediator, finding that superexchange is strongly enhanced as a consequence. In Sec.~\ref{sec:Linear_four_electrons}, we compare the triangular triple-dot case with four electrons to the linear triple-dot case. We present our conclusions in Sec.~\ref{sec:conclusions}. An appendix contains additional details about our calculations and a detailed survey of the single-electron density for each of the triple-dot configurations considered.

\section{Quantum dot model and exchange energy for double dot}\label{sec:CI_two_dots}
We model each quantum dot by a 2d symmetric parabolic potential, and we include a uniform external magnetic field in the $z$ direction, which is orthogonal to the plane of the dots. The Hamiltonian for $N$ electrons is then
\begin{align}\label{eq:general_ham}
H_{i,N}&=\sum_{k=1}^N\bigg[\frac{1}{2m^*}(-i\hbar\nabla_k+\frac{e}{c}\mathbf{A})^2+V_i(\mathbf{r}_k)+g^*\mu_B\mathbf{B}\cdot\mathbf{S}_k\bigg]\nonumber\\&+\sum_{j<k}\frac{e^2}{\kappa|\mathbf{r}_j-\mathbf{r}_k|},
\end{align}
where $m^*=0.067m_e$ is the effective electron mass for GaAs, $m_e$ is the electron mass, $g^*=-0.44$ is the effective Land\'e factor, $\mu_B$ is the Bohr magneton, and the dielectric constant for this material is $\kappa=13.1\epsilon_0$. Throughout this work, we set $\mathbf{B}=B_0\hat z$, where $B_0=0.845$ T. $V_i$ is the total quantum dot potential for case $i$, where $i=1,...,4$ refers to one of the cases shown in Fig.~\ref{fig:1}. 

Let us first consider a system of two small quantum dots without a mediator, as shown in Fig.~\hyperref[fig:1]{\ref*{fig:1}(a)} (case 1). In the following sections, we use this system as a reference against which we compare triple-dot configurations involving a mediator.
The explicit form of the quantum dot potential in Eq.~\eqref{eq:general_ham} for case $i=1$ is
\begin{align}\label{eq:Potential_case1}
V_1(\mathbf{r})&=\frac{1}{2}\Theta(x)m^*\omega_0^2(\mathbf{r}-\mathbf{R}_{D2})^2\nonumber\\&+\frac{1}{2}\Theta(-x)m^*\omega_0^2(\mathbf{r}-\mathbf{R}_{D1})^2.
\end{align}
Here, $\hbar\omega_0$ is the confinement energy for each dot, $\mathbf{r}$ is the electron coordinate, and $r=\sqrt{x^2+y^2}$ the corresponding radius. $\mathbf{R}_{D1}$ and $\mathbf{R}_{D2}$ are the coordinates of the dot centers for D1 and D2, respectively. The confinement energy of each dot is set to $\hbar\omega_0=7.28$ meV, which for $B_0=0.845$ T corresponds to a radius of about $12.5$ nm, and the center-to-center separation between the two dots is $56$ nm. $\Theta(x)$ is the unit step function, which is used to cut and glue the two harmonic oscillator potentials together along the $x=0$ line, as indicated by the blue dashed line in Fig.~\hyperref[fig:1]{\ref*{fig:1}(a)} (case 1). Similar cuts are used in the triple-dot cases as well. This is explained further in the next sections. 

To compute the eigenstates and energies of Eq.~\eqref{eq:general_ham}, we employ CI (e.g., exact diagonalization) following the approach used in our previous work ~\cite{Barnes2011,Deng2018}. Our single-particle basis states are comprised of the Fock-Darwin states for each dot. Because the Fock-Darwin states from different dots are not orthonormal to each other, we use the Cholesky decomposition to obtain linear combinations of them that do form a fully orthonormal basis. After constructing our single-particle basis and truncating it to retain only the lowest $L$ levels, we build our multi-particle states and compute matrix elements of the Hamiltonian with respect to these using the Slater-Condon rules (see the appendix of Ref.~\cite{Barnes2011} for a review). We then extract the effective exchange energy $J$ by computing the energy difference of the lowest-energy triplet $\ket{T_0}$ and singlet $\ket{S}$ states. We establish convergence by adjusting $L$ until the results do not change significantly.

\begin{figure}[!tbp]
\centering
\subfloat[Exchange energy versus number $L$ of single-particle orbitals used in CI calculation for two electrons in a double quantum dot.]
{{\includegraphics[trim=0cm 0cm 0cm 0cm, clip=true,width=8.7cm, angle=0]{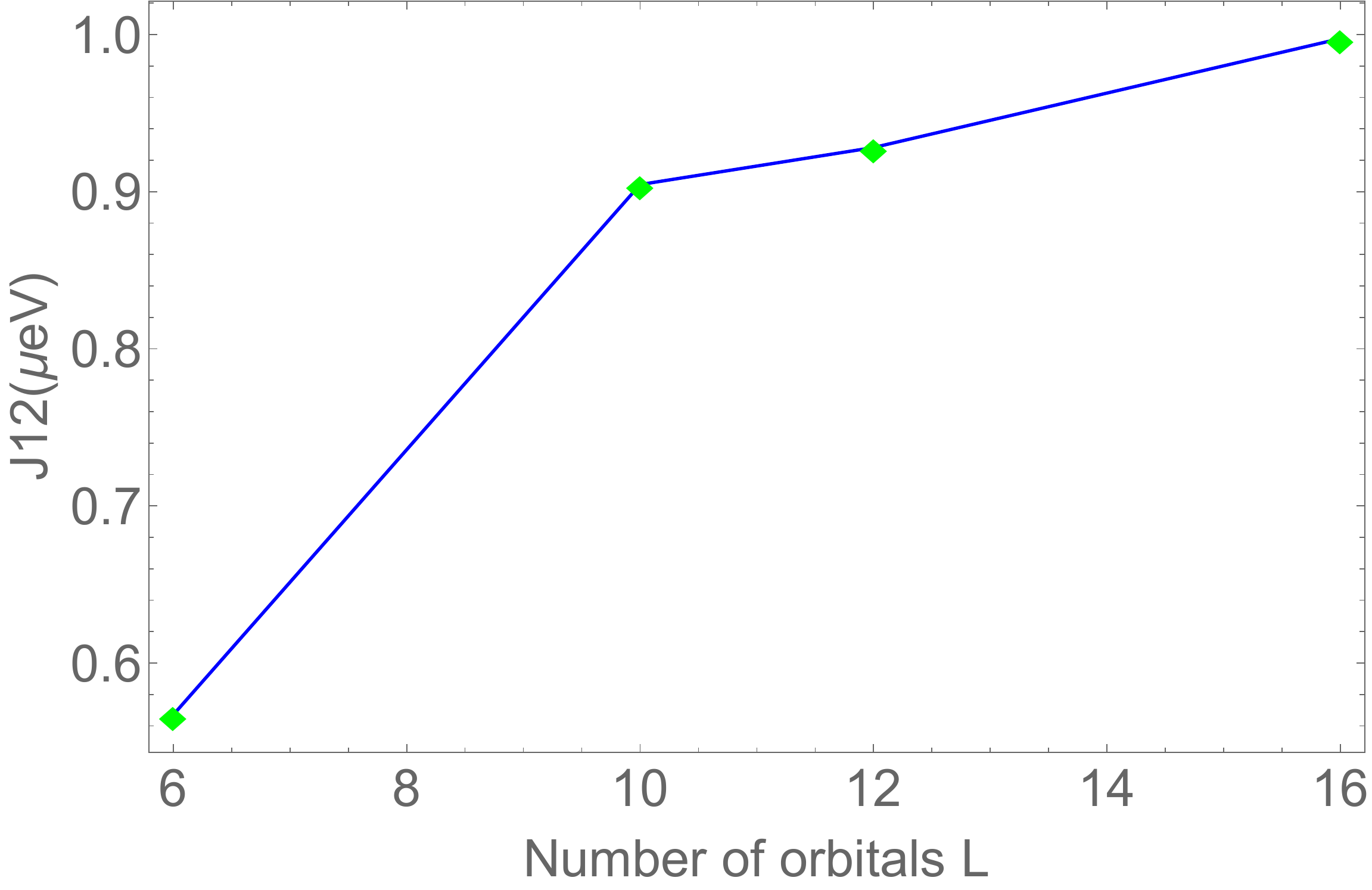}}}\\
\subfloat[Exchange energy versus inter-dot half-distance $l_0$.]
{{\includegraphics[trim=0cm 0cm 0cm 0cm, clip=true,width=8.7cm, angle=0]{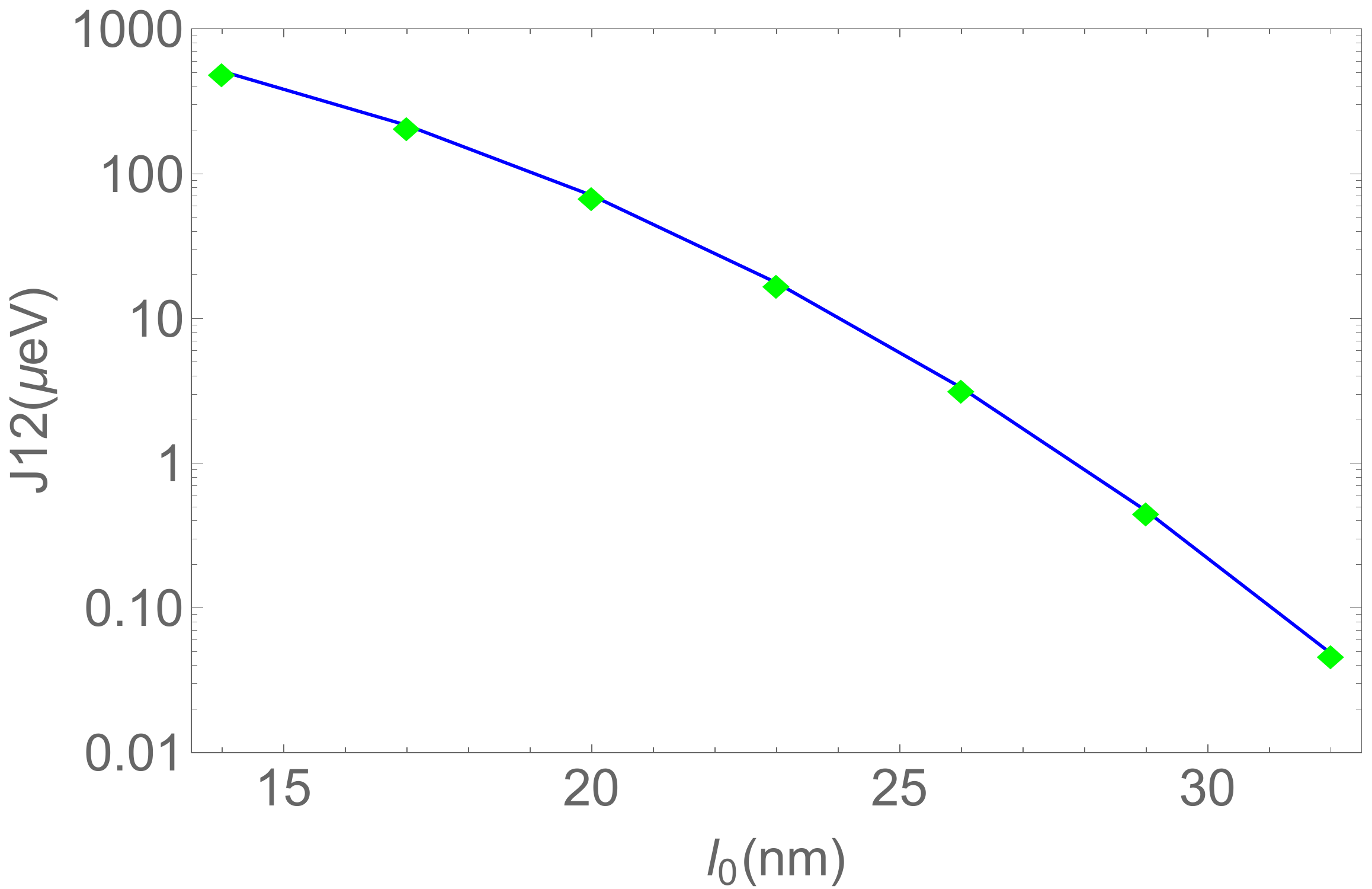}}}
\caption{(a) Exchange energy for two quantum dots with one electron each (see Fig.~\hyperref[fig:1]{\ref*{fig:1}(a)}) computed from CI as a function of the number $L$ of single-particle orbitals ($L/2$ orbitals for each dot). Full orbital shells on each dot are retained when $L=6$ and $L=12$. (b) Exchange energy for two dots versus inter-dot half-distance $l_0$ from a CI calculation with $L=12$ single-particle basis states.}\label{fig:2}
\end{figure}

Fig.~\hyperref[fig:2]{\ref*{fig:2}} shows our CI results for case 1, two electrons in two small dots.  Here in Fig.~\hyperref[fig:2]{\ref*{fig:2}}(a), the orbital number is the total number of orthonormalized Fock-Darwin orbitals for the whole system. The energy levels of the Fock-Darwin states are $E_{n,m}=(n+1)\hbar\sqrt{\omega_0^2+\omega_c^2/4}+m\hbar\omega_c/2$, where $\omega_c=eB_0/m^*c$ is the cyclotron frequency, $n$ is a non-negative integer, and $m=-n,-n+2,...,n-2,n$ is the magnetic quantum number. In the limit of zero magnetic field, these levels form degenerate shells labeled by the quantum number $n$, where the degeneracy of the $n$th shell is $n+1$. For the relatively weak magnetic field considered here, for which $\omega_c=2\omega_0/\sqrt{99}$, the levels in each shell are nearly degenerate. Because of this, one might expect that it is necessary to retain all the orbitals within a shell in order for results to converge. Although Fig.~\hyperref[fig:2]{\ref*{fig:2}}(a) indicates that this may not really be necessary (orbital numbers 6 and 12 correspond to keeping full shells in this example), we choose to retain an integer number of shells in our CI calculations throughout this work to be safe. In Fig.~\hyperref[fig:2]{\ref*{fig:2}}(b), we show the exchange energy as a function of the inter-dot half-distance $l_0$ (measured from the center of D1 to the origin) on a logarithmic scale. As expected, the exchange energy falls of exponentially with the distance. We compare this result to what happens when the mediator dot D3 is placed at the origin in Sec.~\ref{sec:Linear_four_electrons}.

\section{Triangular triple dot with two electrons}\label{sec:Triangular_two_electrons}
We now move on to case 2, which includes a third, larger dot as shown in Fig.~\hyperref[fig:1]{\ref*{fig:1}(b)}. We expect that if the third dot is brought sufficiently close to the first two, then it can mediate superexchange interactions between the electrons on the two small dots. These superexchange interactions can potentially combine constructively or destructively with the normal exchange that still exists between the two small dots.

\begin{figure}[!tbp]
\centering 
\includegraphics[trim=0cm 0cm 0cm 0cm, clip=true,width=8.7cm, angle=0]{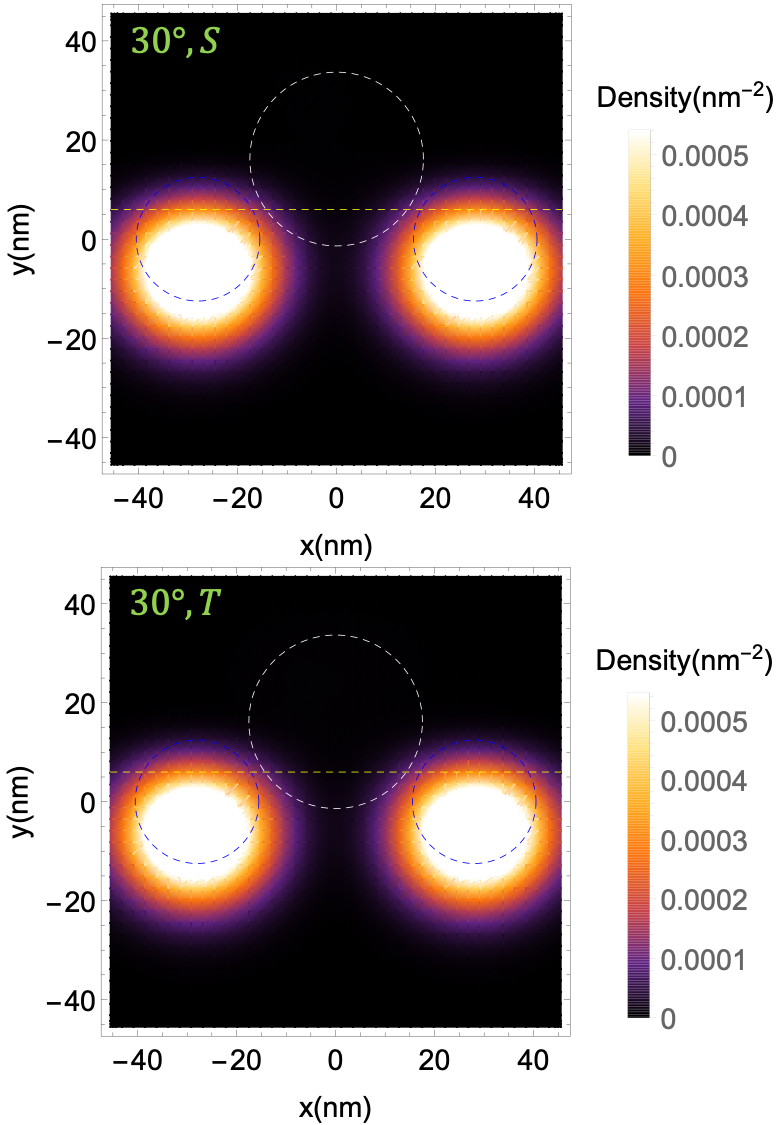}
\caption{Single-particle density for two electrons in a triangular triple dot (case 2) for $\alpha=\ang{30}$. The upper (lower) panel shows the single-particle density for the lowest-energy singlet $\ket{S}$ (triplet $\ket{T_0}$). The yellow dashed line is the potential cut at $y=y_0$. The dashed circles mark the positions of the three parabolic dot potentials for this value of $\alpha$. It is apparent that the centers of the electron density in the small dots are displaced downward. This is due to the large value of detuning $\Delta$ chosen to deplete the big dot. The distance between the two small dots remains the same, as does the normal exchange coupling between them.}\label{fig:3}
\end{figure}

In order to compare directly with the results of case 1, we again fix the radius of the two small dots (D1 and D2) to about $12.5$ nm and choose the center-to-center distance ($2l_0$) between them to be $56$ nm. The larger mediator dot (D3) is chosen to have a confinement energy of $\hbar\omega_1=\frac{1}{2}\hbar\omega_0=3.64$ meV, which corresponds to a radius of about $17.5$ nm. The center of the dot is located on the positive $y$ axis. The centers of these three dots form an isosceles triangle, and the angle between the base and one leg of the triangle is defined to be $\alpha$. We match the three parabolic potentials of the three dots along a T cut that separates the plane into three regions as indicated with the blue dashed lines in Fig.~\hyperref[fig:1]{\ref*{fig:1}(b)}. Each of these regions contains one of the dot potentials. The horizontal separation line is placed at $y=y_0$, where $y_0$ depends on the angle $\alpha$ in such a way that dot D3 remains almost entirely above this line in all cases. The precise manner in which $y_0$ is chosen for a given value of $\alpha$ is described in Appendix ~\ref{app:A}. The remaining two regions are separated by the $y$-axis at $x=0$. Thus, for case $i=2$, the total quantum dot potential in  Eq.~\eqref{eq:general_ham} is
\begin{align}\label{eq:Potential_case2}
V_2(\mathbf{r})=&\frac{1}{2}\Theta(y_0-y)\Theta(x)m^*\omega_0^2(\mathbf{r}-\mathbf{R}_{D2})^2\nonumber\\
&+\frac{1}{2}\Theta(y_0-y)\Theta(-x)m^*\omega_0^2(\mathbf{r}-\mathbf{R}_{D1})^2\nonumber\\
&+\Theta(y-y_0)[\frac{1}{2}m^*\omega_1^2(\mathbf{r}-\mathbf{R}_{D3})^2+\Delta].
\end{align}

\begin{figure}[!tbp]
\centering 
\includegraphics[trim=0cm 0cm 0cm 0cm, clip=true,width=8.7cm, angle=0]{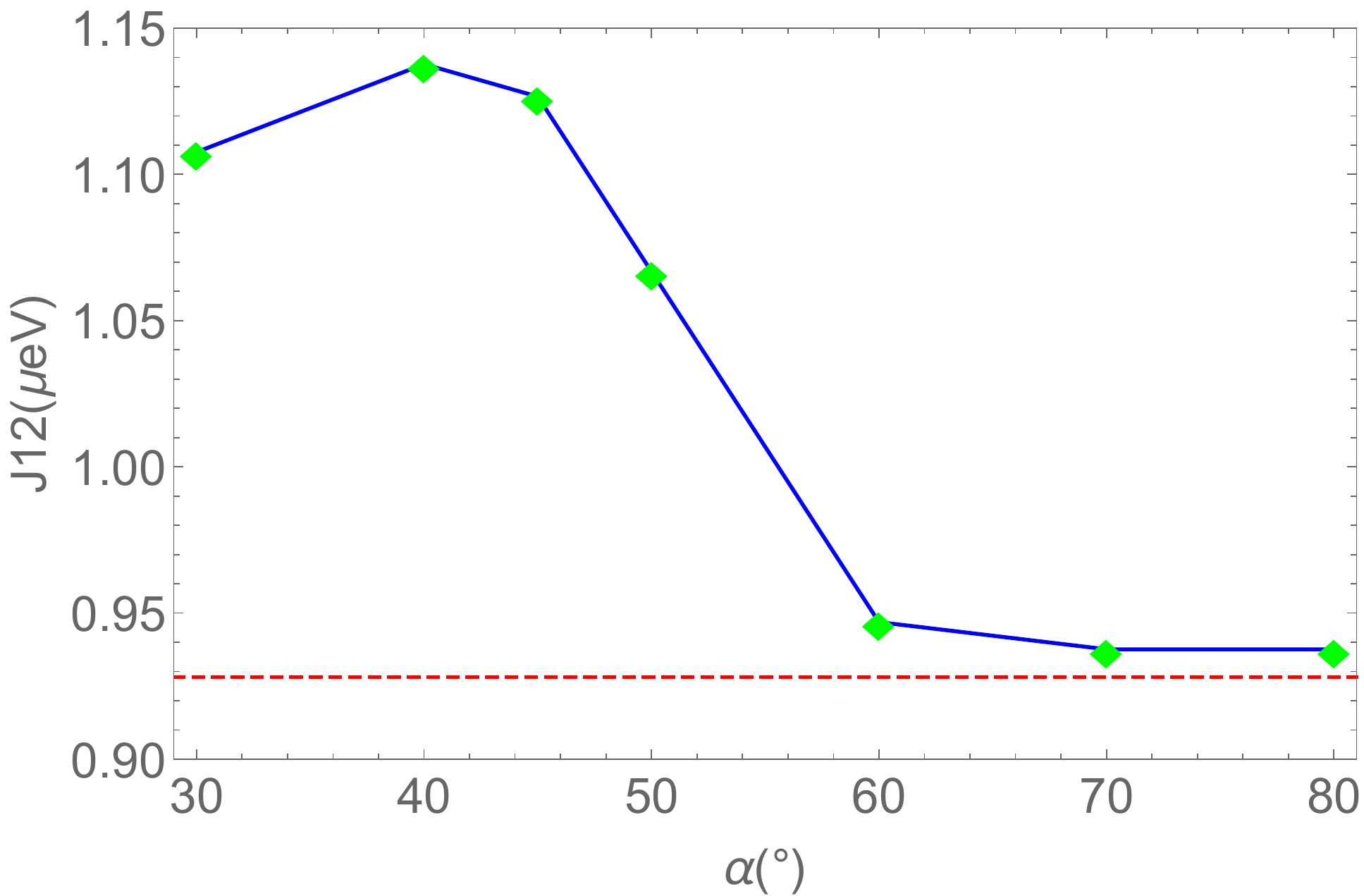}
\caption{Effective exchange interaction between electrons on dots D1 and D2 as a function of angle $\alpha$ (see Fig.~\hyperref[fig:1]{\ref*{fig:1}(b)}) in the case of two electrons in a triangular triple dot (case 2). In the CI calculation used to obtain this result, we retain 6, 6 and 3 orthonormalized Fock-Darwin orbitals for D1, D2 and D3, respectively, corresponding to a total of $L=15$ single-particle basis states. The red dashed line is the result for $J_{12}$ from case 1 (two electrons in a double dot) using 12 single-particle orbitals. We see only modest contributions from superexchange processes in this case.}\label{fig:4}
\end{figure}

In the last line of Eq.~\eqref{eq:Potential_case2}, we introduced the detuning parameter $\Delta$ for dot D3; this parameter applies a constant energy shift to all the levels in the mediator relative to the energy levels in the small dots, D1 and D2. For each angle $\alpha$, we adjust $\Delta$ until dot D3 is empty of electrons, which we check by integrating the multi-particle density over the region above the horizontal blue dashed line in Fig.~\hyperref[fig:1]{\ref*{fig:1}(b)} to obtain the electron number in D3. Of course, one cannot make the electron number in D3 exactly zero, but it can be made very small, at least for $\alpha\geq\ang{45}$. For these angles, we can keep the electron number in D3 below $0.05e$ by choosing $\Delta=20$ meV for all angles in this range. For smaller angles in the range $\alpha<\ang{45}$, reducing the electron number in D3 becomes more difficult, but we can still get it below $0.1e$ by setting $\Delta=20$ meV.

The successful depletion of D3 is also visible in the single-particle density, which we calculate by integrating the multi-electron density over just one set of electronic coordinates. An example is shown in Fig.~\hyperref[fig:3]{\ref*{fig:3}} for $\alpha=\ang{30}$. Increasing $\Delta$ until the big dot is almost vacant also causes the two small dots to move downward, as is evident in the figure. However, the separation between the electrons in the small dots remains unchanged for both the lowest-energy singlet and triplet state, hence the normal exchange interaction between them remains the same. Additional density plots for various angles can be found in Appendix~\ref{app:A}.

We now calculate the effective exchange interaction $J_{12}$ between the electrons on dots D1 and D2. This interaction includes contributions from normal, nearest-neighbor exchange between D1 and D2 as well as a superexchange interaction mediated by dot D3. We compute $J_{12}$ as a function of angle $\alpha$; this allows us to control the relative strength of the superexchange coupling compared to the normal exchange, because increasing $\alpha$ increases the distance between $D3$ and the other two dots. The result is shown in Fig.~\hyperref[fig:6]{\ref*{fig:6}} (blue line with green points). We see that across a broad range of angles, the effective $J_{12}$ exceeds the normal exchange interaction (dashed red line) that we obtain in the absence of the mediator. We attribute the difference between these two curves to superexchange processes. It is evident that, in this case, superexchange provides only a modest enhancement of the total effective exchange that is at most 20\% of the normal exchange. This enhancement quickly fades as $\alpha$ increases beyond $\ang{50}$ (which corresponds to $l_1=43.56$ nm), although some evidence of superexchange remains visible in the large-angle regime. Interestingly, we also find non-monotonic behavior in $J_{12}$ in the small angle regime, with a maximum near $\ang{40}$. This may be because the downward shift of the electrons in D1 and D2 caused by $\Delta$ effectively increases the distance to D3, leading to a small suppression of superexchange.

\section{Triangular triple dot with four electrons}\label{sec:Triangular_four_electrons}
We now investigate the impact of increasing the number of electrons on the effective exchange coupling. In particular, we add two more electrons to the system, while keeping the form of the potential ($V_3=V_2$) and almost all the parameters the same. The only parameter we change is $\Delta$, which is now adjusted so that two electrons occupy D3 as in Fig.~\hyperref[fig:1]{\ref*{fig:1}(c)}. The precise values used are given in Appendix~\ref{app:A}. 

\begin{figure}[!tbp]
\centering 
\includegraphics[trim=0cm 0cm 0cm 0cm, clip=true,width=8.7cm, angle=0]{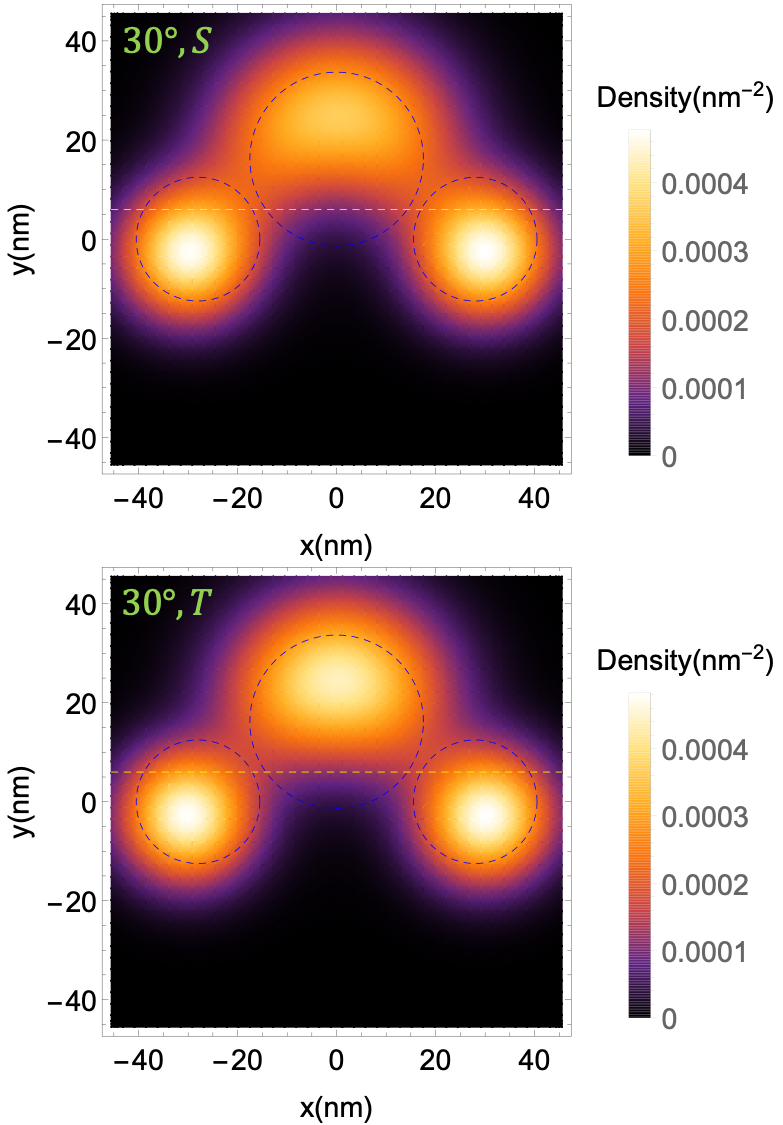}
\caption{Single-particle density for four electrons in a triangular triple dot (case 3) with $\alpha=\ang{30}$. The upper (lower) panel shows the single-particle density for the lowest-energy singlet-like $\ket{S}$ (triplet-like $\ket{T_0}$) four-electron state. The yellow dashed line marks the potential cut at $y=y_0$. The blue dashed circles indicate the original positions of the three dots for this value of $\alpha$. The electrons in the small dots are displaced slightly downward and outward, while the electrons in the big dot move upward due to Coulomb repulsion.}\label{fig:5}
\end{figure}

Before proceeding, we need to clarify the definition of $J_{12}$ in the case where there are four electrons. For each possible occupancy of spatial orbitals, there are a total of 16 spin states, many of which are singlet-like and triplet-like. To compute the effective exchange energy, we identify the lowest-energy state with $S^{total}=0$ and $S^{total}_{z}=0$ as our singlet state $\ket{S}$ and the lowest-energy state with $S^{total}=1$ and $S^{total}_{z}=0$ as our triplet state $\ket{T_0}$. We then calculate $J_{12}$ by taking the difference of the two corresponding eigenenergies. In all four-electron cases considered in this work, we confirm the suitability of this definition by verifying that the resulting $\ket{S}$ and $\ket{T_0}$ states have the property that the two electrons on D3 approximately form a singlet.

The one-electron density (obtained this time by integrating the full multielectron density over three sets of electronic coordinates) for $\alpha=\ang{30}$ is shown for both the lowest-energy singlet-like and triplet-like states in Fig.~\hyperref[fig:5]{\ref*{fig:5}}. It is evident in both cases that the electrons in D1 and D2 are displaced slightly downward as a consequence of D3, similarly to Fig.~\hyperref[fig:3]{\ref*{fig:3}}, while the two electrons on D3 undergo a more substantial upward shift. This is due to a combination of Coulomb repulsion and the fact that the confinement energy of D3 is much smaller than that of D1 and D2, which allows the electrons in D3 more freedom to move away from the other two electrons. Unlike the two-electron case, here the electrons in D1 and D2 are also pushed away from each other horizontally as a consequence of the Coulomb repulsion from the two electrons on D3, and this in turn can impact the normal exchange. As one would anticipate based on fermion statistics, it is also evident in Fig.~\ref{fig:5} that the singlet density is more uniformly spread across the three dots compared to the triplet density. This is, of course, directly related to the nonzero superexchange energy. Additional plots of the single-particle density for other values of $\alpha$ can be found in Appendix~\ref{app:A}. 

\begin{figure}[!tbp]
\centering
\subfloat[Effective exchange energy for triangular triple dot with four electrons.]
{{\includegraphics[trim=0cm 0cm 0cm 0cm, clip=true,width=8.7cm, angle=0]{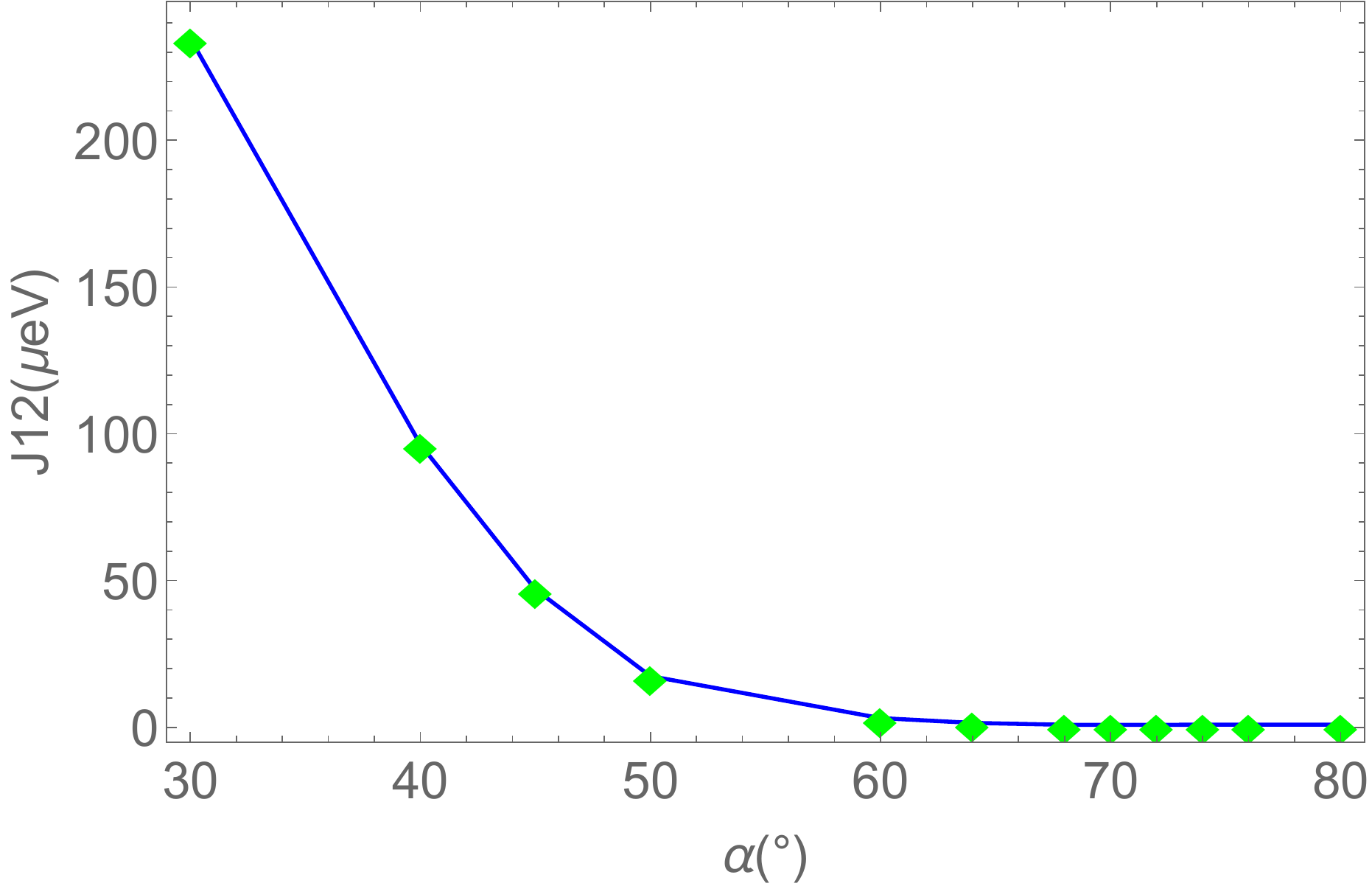}}}\\
\subfloat[Zoom-in of panel (a).]
{{\includegraphics[trim=0cm 0cm 0cm 0cm, clip=true,width=8.7cm, angle=0]{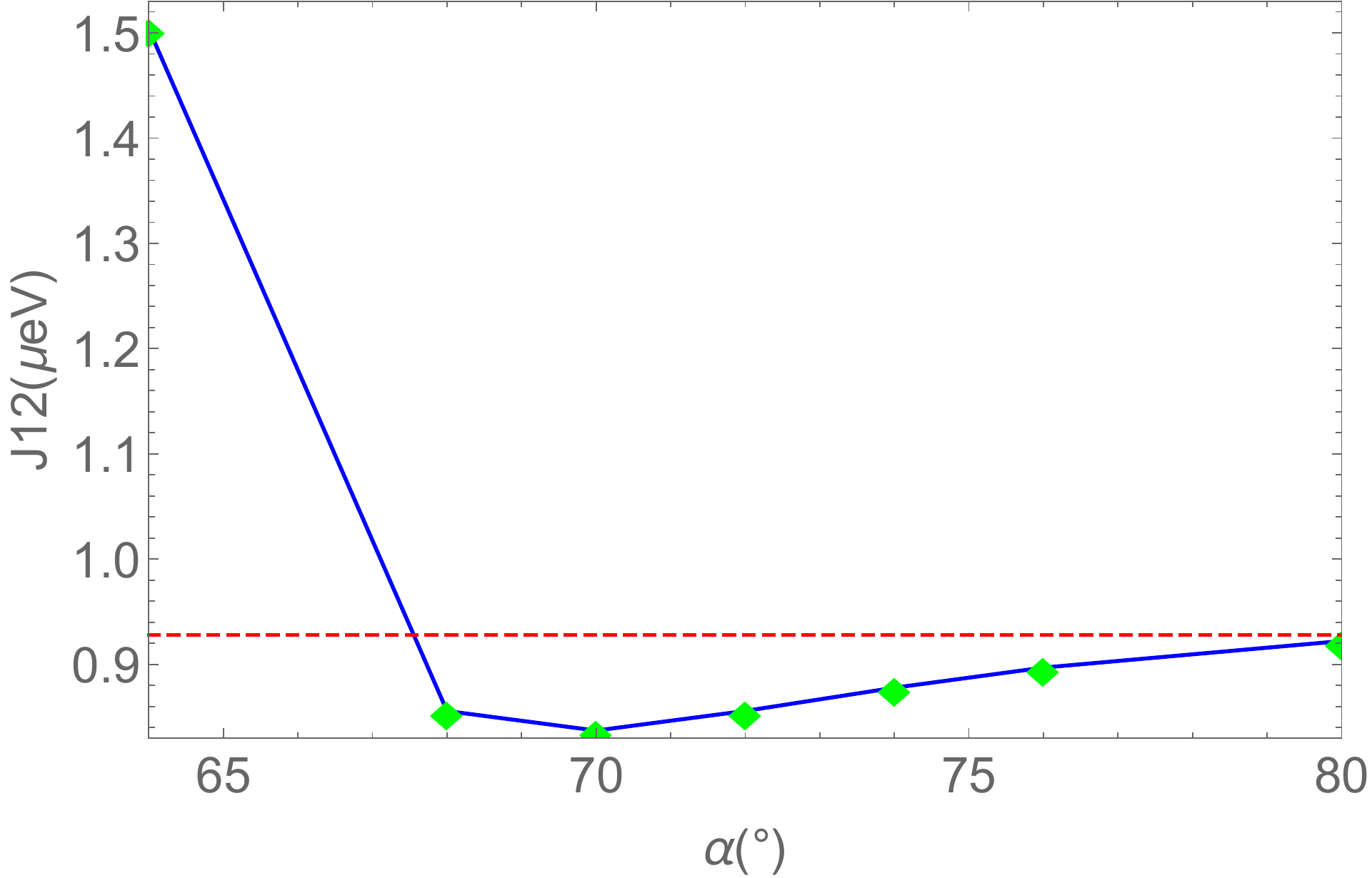}}}
\caption{(a) Effective exchange energy $J_{12}$ versus angle $\alpha$ for four electrons in a triple quantum dot as in Fig.~\hyperref[fig:1]{\ref*{fig:1}(c)} (case 3). Results are obtained from a CI calculation in which 6, 6 and 3 orbitals are retained for D1, D2 and D3, respectively, for a total of $L=15$ single-particle basis states. Superexchange processes strongly enhance $J_{12}$ for $\alpha<\ang{65}$. (b) A zoom-in of (a), along with the CI result for two electrons in a double dot (case 1) when $L=12$ single-particle states are kept (red dashed line). The dip in $J_{12}$ near $\alpha=\ang{70}$ is likely caused by Coulomb repulsion as explained in the text.}\label{fig:6}
\end{figure}

Next, we show that the inclusion of the two additional electrons compared to case 2 can either strongly enhance or weakly suppress the effective exchange interaction $J_{12}$ depending on $\alpha$. Our CI results for $J_{12}$ as a function of $\alpha$ for case 3 are shown in Fig.~\hyperref[fig:6]{\ref*{fig:6}}. The most striking difference compared to the two-electron case considered in the previous section is that the effective exchange energy is more than two orders of magnitude larger in the low-angle regime (compare with Fig.~\hyperref[fig:4]{\ref*{fig:4}}). We can understand this as a consequence of Fermi statistics combined with the fact that the three quantum dot potentials are merging together in the low-$\alpha$ regime, which forces the four electrons to occupy the same space. This happens in the four-electron case because here we lower $\Delta$ substantially in order to keep two electrons trapped in the mediator. This is unlike the previous case where $\Delta$ was set to a large value to keep the mediator empty, which in turn keeps the potential similar to what it was in the case of two isolated dots (case 1). If we were to think of the triple-dot potential in case 3 (low $\Delta$) as effectively one big dot D$_{eff}$, and if we neglect Coulomb interactions for the moment, then the ground state of the system would be a $S^{total}=0$ state consisting of two pairs of two-electron singlets occupying the lowest two single-particle orbitals of D$_{eff}$. The lowest-energy state with $S^{total}=1$ and $S_z^{total}=0$ would be formed by moving one of the electrons to the second excited orbital, which produces an exchange splitting that is on the order of the level spacing of D$_{eff}$. In the present context, this is on the order of meV. (In case 2 where $\Delta$ is large, this splitting is very small because we have essentially two independent dots with nearly degenerate orbitals.) Restoring the Coulomb interactions reduces this splitting because the more symmetric spatial part of the $S^{total}=0$ state incurs a larger Coulomb energy penalty, but the splitting can still remain large. Note that this mechanism is closely related to the notion of spin blockade in singlet-triplet qubits, where the singlet and triplet two-electron states are nearly degenerate when the electrons are separated into distinct dots, but when a large detuning is applied to one dot, the electrons are pushed into the same dot, opening a large energy gap between the singlet and triplet states \cite{Petta2005}. In the next section, we show that the vertical shift of the mediator electrons due to Coulomb repulsion (Fig.~\ref{fig:5}) actually leads to a significant reduction in the superexchange in the small-angle regime compared to what would occur in the absence of this shift.

Fig.~\hyperref[fig:6]{\ref*{fig:6}} shows that the effective exchange energy exhibits three different qualitative trends as a function of $\alpha$. For $\alpha<\ang{65}$, $J_{12}$ is dominated by a very strong superexchange interaction mediated by D3 as discussed above. For $\alpha>\ang{72}$, $J_{12}$ quickly converges to the value obtained in the two-electron case without the mediator (which is indicated with a red dashed line in the figure). In between these regimes, $\ang{68}<\alpha<\ang{72}$, $J_{12}$ is close to but clearly below the two-dot value. One possible explanation for this behavior is that the superexchange contribution is becoming negative in this range and partially cancels the positive normal exchange energy. Negative superexchange couplings mediated by large quantum dots have recently been observed experimentally \cite{Malinowski2019}. We have also shown in prior work that negative exchange can arise in quantum dots containing as few as four electrons \cite{Deng2018}. However, we believe that it is more likely that the superexchange coupling quickly drops to zero before $\alpha=\ang{68}$, and that the suppression of $J_{12}$ after this point is instead due to the horizontal displacement of the electrons in D1 and D2 caused by Coulomb repulsion as shown in Fig.~\ref{fig:5}. We have checked numerically that displacing dots D1 and D2 by a similar amount in the two-dot geometry leads to a change in $J_{12}$ that is of the same order of magnitude in that case, supporting this interpretation. This effect could have important consequences in general for architectures in which long-distance interactions are mediated by multielectron quantum dots because, in addition to mediating superexchange interactions, the extra electrons on the mediators can also have a negative impact on the resulting spin-spin coupling strength due to Coulomb interactions depending on the layout of the dots.

\section{Linear triple dot with four electrons}\label{sec:Linear_four_electrons}
We now move on to the final quantum dot configuration considered in this work: the linear triple dot with four electrons depicted in Fig.~\hyperref[fig:1]{\ref*{fig:1}(d)}.  The total quantum dot potential in this case is
\begin{align}\label{eq:Potential_case1}
V_4=&\frac{1}{2}\Theta(x-x_0)m^*\omega_0^2(\mathbf{r}-\mathbf{R}_{D2})^2\nonumber\\
&+\frac{1}{2}\Theta(-x_0-x)m^*\omega_0^2(\mathbf{r}-\mathbf{R}_{D1})^2\nonumber\\
&+\Theta(x+x_0)\Theta(x_0-x)[\frac{1}{2}m^*\omega_1^2(\mathbf{r}-\mathbf{R}_{D3})^2+\Delta].
\end{align}
This potential is formed by cutting and gluing together the individual dot potentials along vertical lines located at $x=\pm x_0$. We are interested in computing the effective exchange energy between dots D1 and D2 as a function of the inter-dot distance $l_1$. For each $l_1$, we set the detuning $\Delta$ such that the mediator is occupied by two electrons. The particular values used are given in Appendix~\ref{app:A}.

\begin{figure}[!tbp]
\centering 
\includegraphics[trim=0cm 0cm 0cm 0cm, clip=true,width=8.7cm, angle=0]{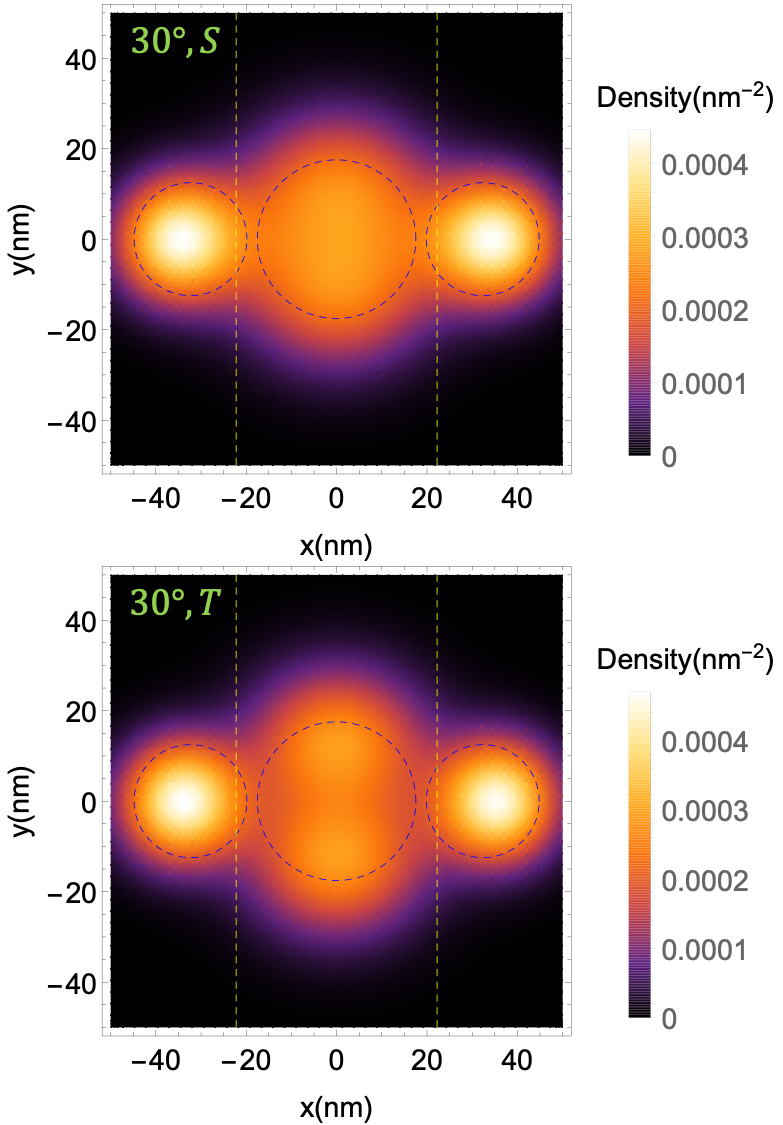}
\caption{Single-particle density for four electrons in a linear triple dot (case 4) with $\alpha=\ang{30}$. The upper (lower) panel shows the single-particle density for the lowest-energy singlet-like $\ket{S}$ (triplet-like $\ket{T_0}$) four-electron state. The yellow dashed lines mark the potential cuts at $x=\pm x_0$. The blue dashed circles indicate the positions of the three dots for this value of $\alpha$. The electrons in the small dots are displaced slightly outward due to Coulomb repulsion.}\label{fig:7}
\end{figure}

\begin{figure}[!tbp]
\centering
\subfloat[Effective exchange energy for linear triple dot with four electrons.]
{{\includegraphics[trim=0cm 0cm 0cm 0cm, clip=true,width=8.7cm, angle=0]{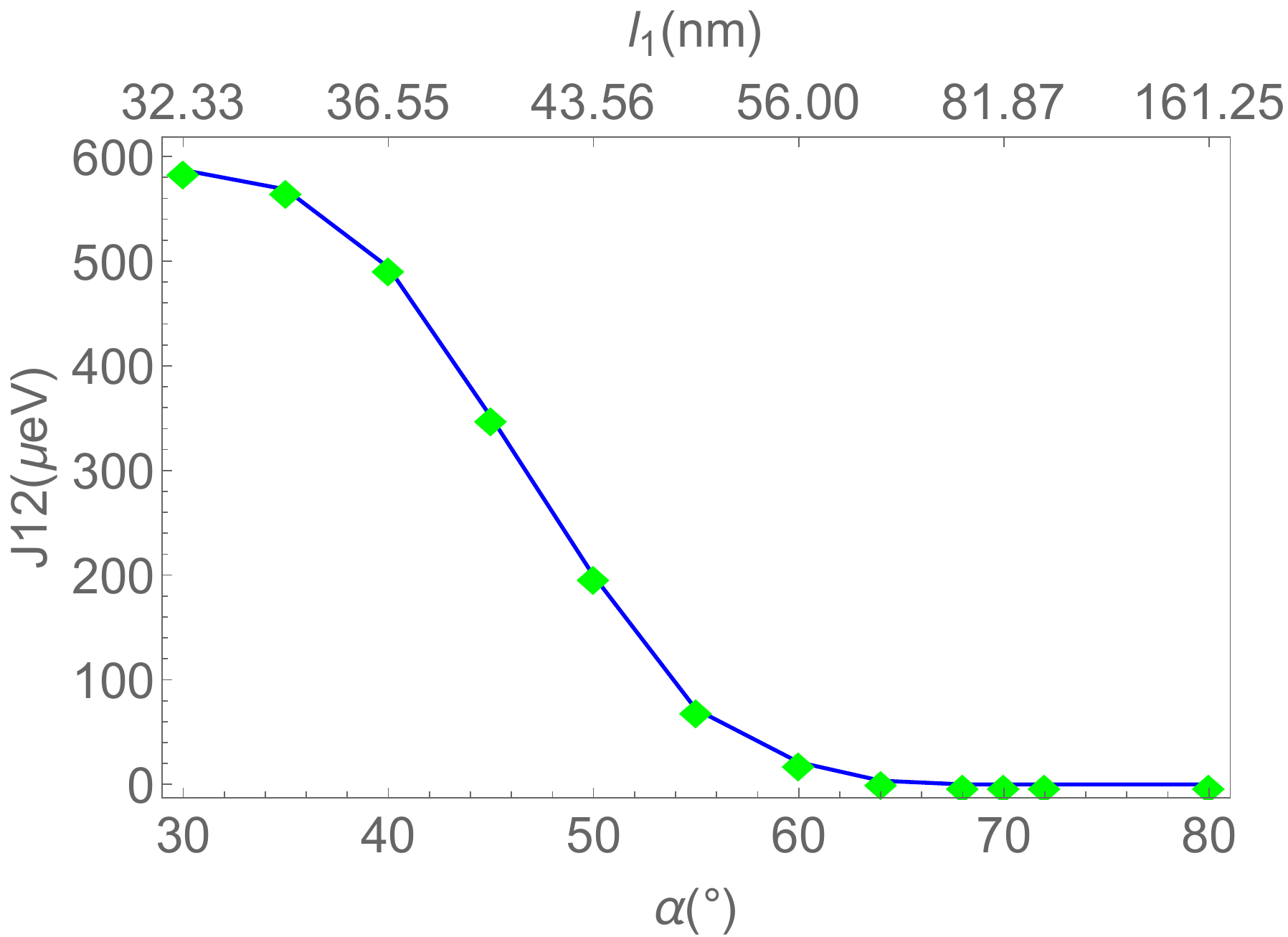}}}\\
\subfloat[Zoom-in of panel (a).]
{{\includegraphics[trim=0cm 0cm 0cm 0cm, clip=true,width=8.7cm, angle=0]{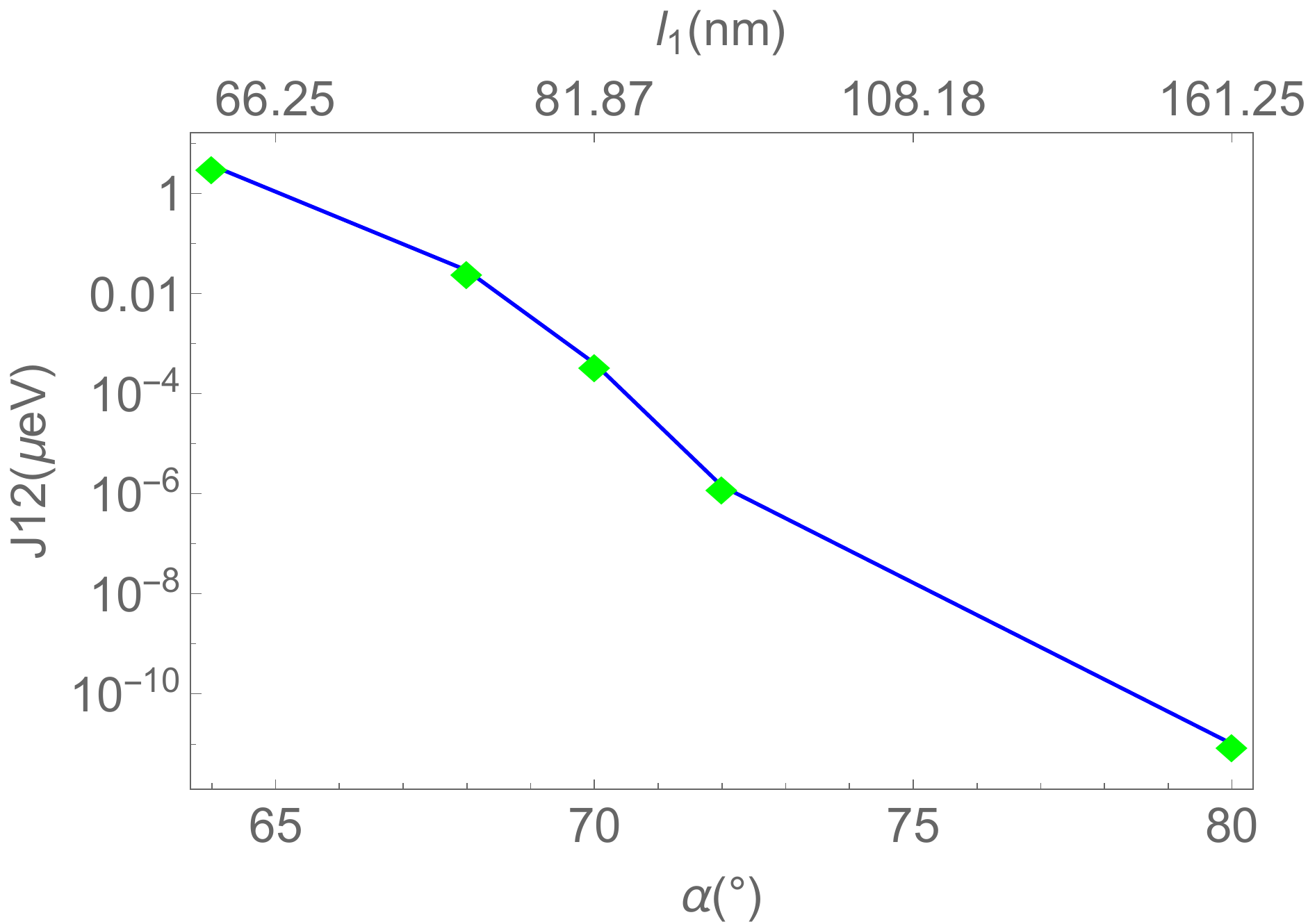}}}
\caption{(a) Effective exchange energy $J_{12}$ versus inter-dot separation (given by $2l_1=2l_0\sec\alpha$, with $l_0=28$ nm) for four electrons in a linear triple dot as in Fig.~\hyperref[fig:1]{\ref*{fig:1}(d)} (case 4). Results are obtained from a CI calculation in which 6, 6 and 3 orbitals are retained for dots D1, D2 and D3, respectively, for a total of $L=15$ single-particle basis states. The presence of the two mediator electrons strongly enhances $J_{12}$ for $\alpha<\ang{65}$. (b) A zoom-in of (a). In the large-distance regime (large $\alpha$ and $l_1$), $J_{12}$ monotonically approaches zero.}\label{fig:8}
\end{figure}

\begin{figure*}[!tbp]
\centering
\subfloat[Interpolating function for effective exchange in case 4.]
{{\includegraphics[trim=0cm 0cm 0cm 0cm, clip=true,width=6.1cm, angle=0]{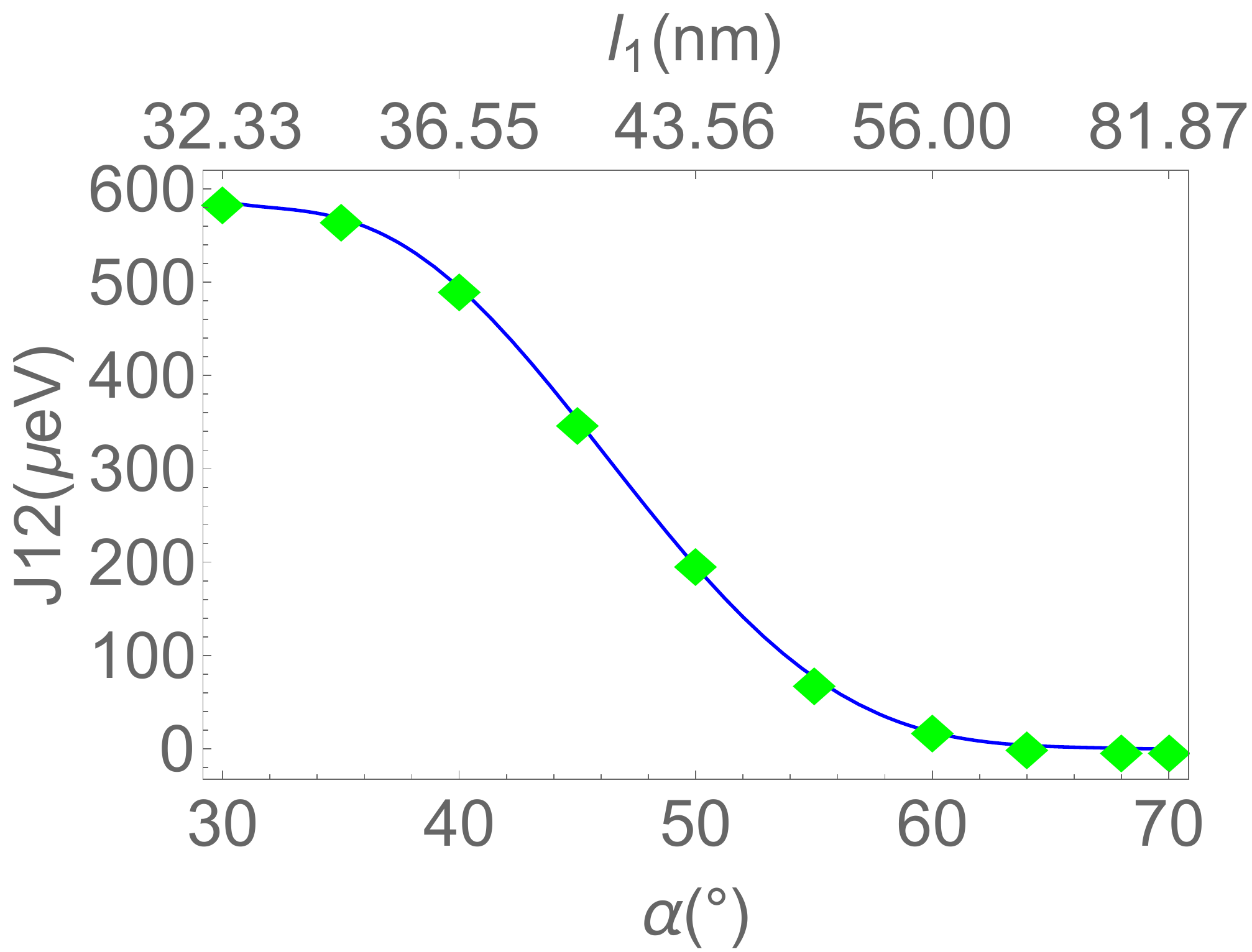}}}
\subfloat[Effective angular shifts for case 3.]
{{\includegraphics[trim=0cm 0cm 0cm 0cm, clip=true,width=5.9cm, angle=0]{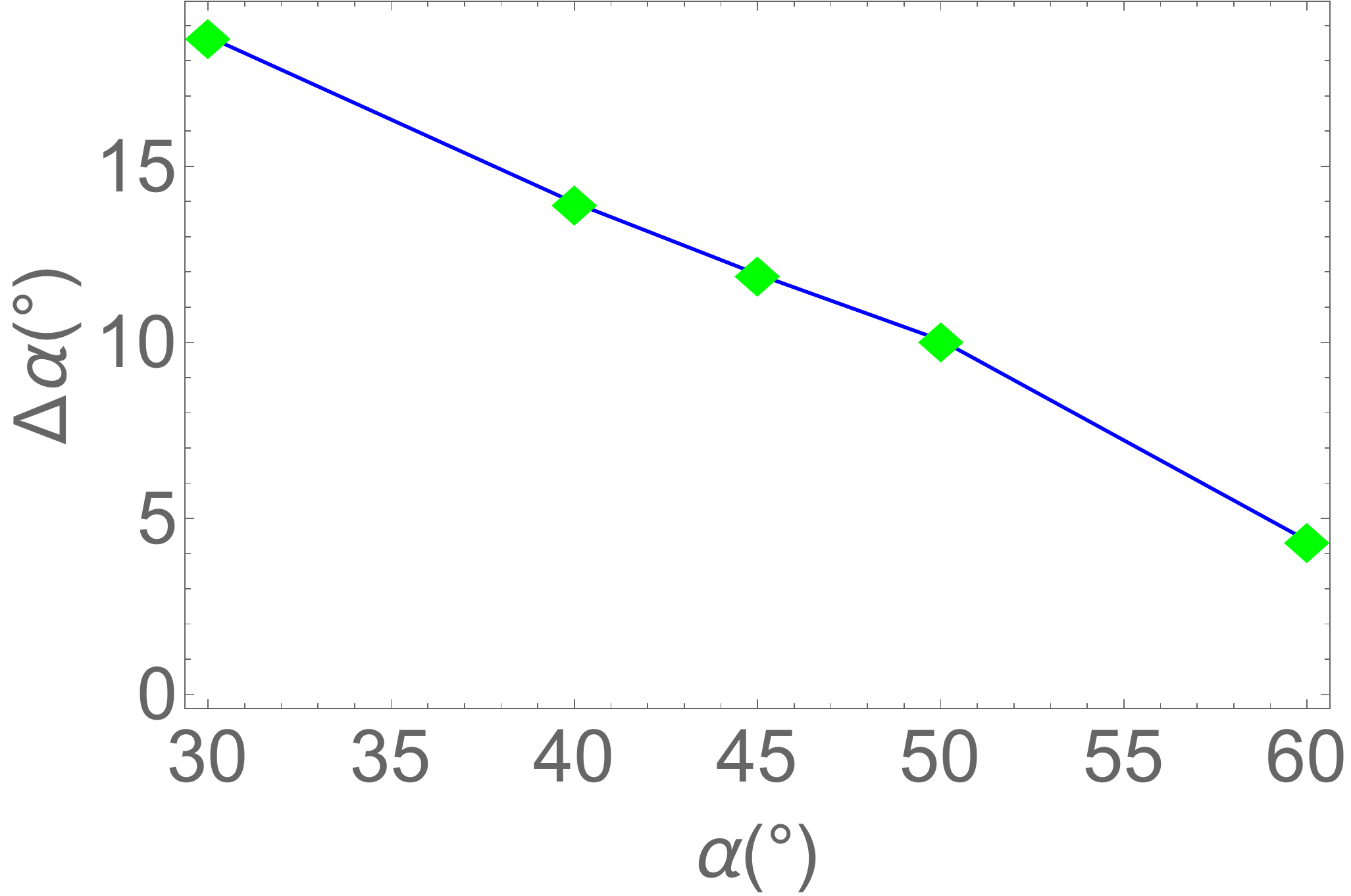}}}
\subfloat[Relative electron position shifts for case 3.]
{{\includegraphics[trim=0cm 0cm 0cm 0cm, clip=true,width=5.9cm, angle=0]{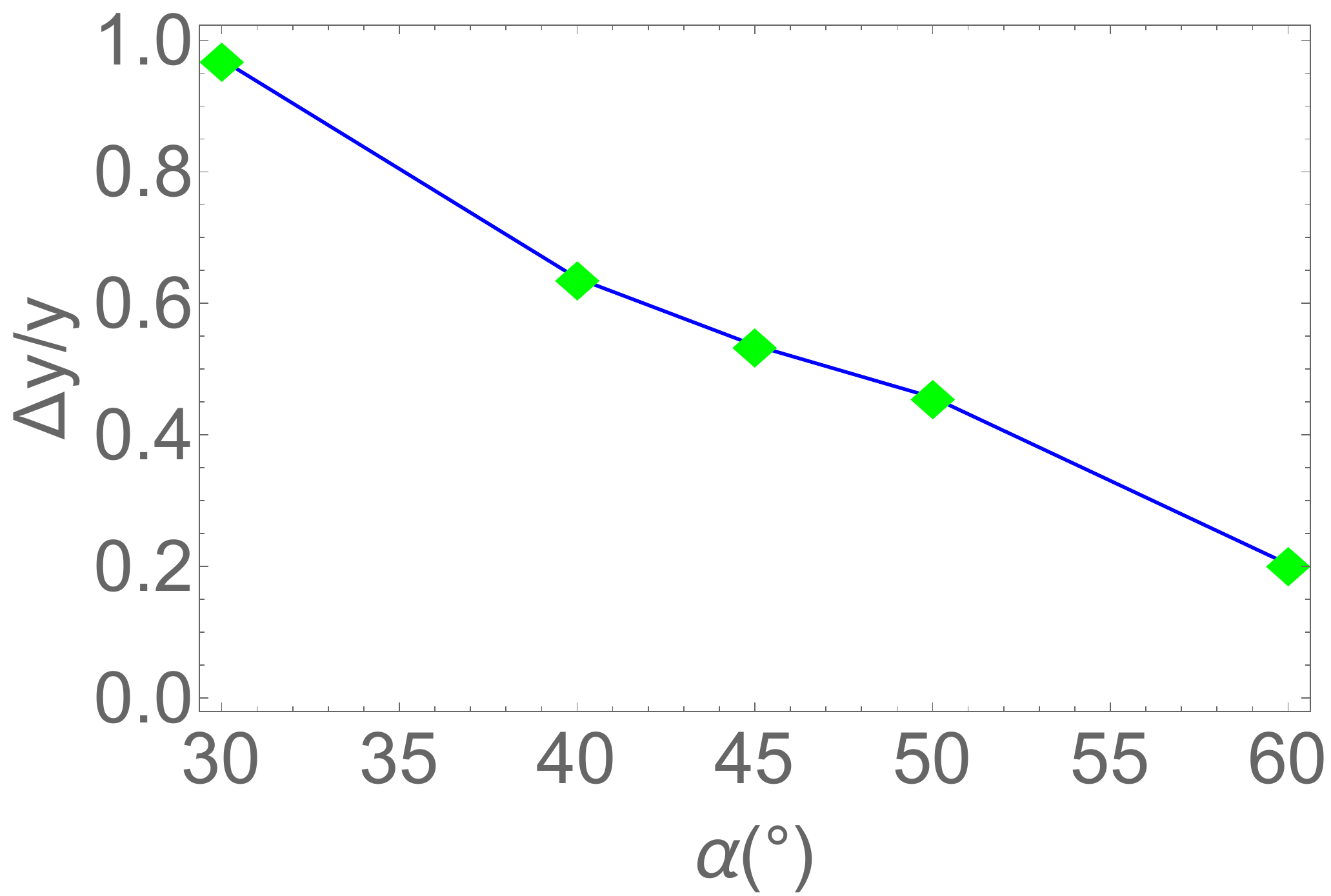}}}
\caption{Calculation of the vertical shift in electron positions due to Coulomb repulsion in case 3 (triangular triple dot) using the exchange energy from case 4 (linear triple dot). We assume the electrons in D1 and D2 remain fixed and only consider the movement of electrons in D3. (a) Effective exchange energy versus $\alpha$ and $l_1$ for case 4. Green diamonds are the original data from Fig.~\ref{fig:8}. The blue line is an interpolating function for lower angles to intermediate angles. One can map the data of case 3 to this function to get information about the shifts in electron positions as explained in the main text. (b) The resulting angular shifts due to Coulomb repulsion for different angles $\alpha$. (c) The relative change in vertical position of the electrons in D3 for different angles.}\label{fig:9}
\end{figure*}

To make it easier to compare directly to the triangular triple dot studied in the previous section (case 3), we again compute $J_{12}$ as a function of angle $\alpha$, but where $\alpha$ now refers to the corresponding angle in the triangular triple dot geometry that has the same distance $l_1$ between the mediator D3 and the smaller dots D1 and D2. Imagine that each linear geometry we consider in this section is obtained by starting from a triangular configuration with angle $\alpha$, freezing $l_1$, and then rotating D1 and D2 around D3 until all three dot centers lie on a line. In this way, each value of $\alpha$ that we start with corresponds to a different inter-dot distance in the linear geometry according to the formula $l_1=l_0\sec\alpha$, where $l_0=28$ nm. We are thus computing $J_{12}(l_1)=J_{12}(l_0\sec\alpha)$ as a function of $\alpha$.

An example of the one-electron density for $\alpha=\ang{30}$ is shown in Fig.~\hyperref[fig:7]{\ref*{fig:7}}. Although the inter-dot distance is rather short at this value of $\alpha$ ($l_1=32.33$ nm), the electrons in D1 and D2 shift only slightly away from D3. This is because of a trade-off between the Coulomb repulsion and the large confinement energy of the small dots. The electrons in the small dots needs to move only a little bit to cancel the Coulomb repulsion from the mediator electrons. The horizontal shift of the D1, D2 electrons is smaller than in case 3 (see Fig.~\ref{fig:5}) because D1 and D2 are a bit further apart here. It is also evident in Fig.~\ref{fig:7} that the density for the triplet state is more disjointed (exhibiting four distinct maxima) compared to the singlet density as one would expect based on fermion statistics. Additional single-electron density plots for other values of $\alpha$ can be found in Appendix~\ref{app:A}.

Our CI results for $J_{12}$ as a function of $\alpha$ in the linear geometry are shown in Fig.~\hyperref[fig:8]{\ref*{fig:8}}. The first thing to notice is that the superexchange again dominates in the small $\alpha$ regime, even more so here than in the triangular triple dot (case 3), as is evident in Fig.~\ref{fig:8}(a). We also see from Fig.~\ref{fig:8}(b) that the interaction strength remains above 1 $\mu$eV for inter-dot distances of up to $\sim 130$ nm. This should be compared to the case without the mediator (Fig.~\ref{fig:2}(b)), where the corresponding distance is only 56 nm. Thus, the mediator extends the interaction range by more than a factor of 2. Fig.~\ref{fig:8}(b) further reveals that the effective exchange decays to zero monotonically with inter-dot distance. Because the horizontal displacement caused by Coulomb repulsion is smaller in this case, a dip in $J_{12}$ does not arise as in case 3. 

The large enhancement in $J_{12}$ at smaller $\alpha$ relative to that seen in case 3 is likely due to the fact that the vertical Coulomb repulsion in case 3 effectively increases the distance to the mediator electrons (see Fig.~\ref{fig:5}), which in turn reduces the benefit to the superexchange that comes from the presence of these extra electrons. The vertical shift is significant because of the weak confinement energy of the mediator. On the other hand, the additional symmetry in the linear geometry prevents a similar phenomenon from happening in this case, yielding a stronger superexchange enhancement. If this interpretation is correct, then we should adjust the $\alpha$ values in case 3 to account for the upward shift of the mediator electrons in order to do a proper comparison to the linear configuration. To check whether this makes sense, we will in fact do the opposite: We first determine the shift in $\alpha^{case3}$ needed to make $J_{12}^{case3}$ equal to $J_{12}^{case4}$ (setting the inter-dot distances equal in both cases), and we will then check whether this shift corresponds to a vertical displacement $\Delta y$ of the mediator electrons that is comparable to that seen in Fig.~\ref{fig:5}. The first step then is to solve the equation
\begin{equation}\label{eq:mapping_angle}
J^{case 3}_{12}(\alpha^{case 3})=J^{case 4}_{12}(\alpha^{case 4}),
\end{equation}
for $\alpha^{case4}$, which we then interpret as the effective angle for case 3: $\alpha_{eff}^{case3}=\alpha^{case4}$. The shift in $\alpha$ caused by Coulomb repulsion is then $\Delta\alpha=\alpha_{eff}^{case3}-\alpha^{case3}$. In order to solve Eq.~\eqref{eq:mapping_angle}, we must first interpolate our data for $J_{12}$ versus $\alpha$ to obtain a smooth function $J_{12}^{case4}(\alpha^{case4})$. This interpolation is shown in Fig.~\ref{fig:9}(a), and the $\Delta\alpha$ that results from solving Eq.~\eqref{eq:mapping_angle} is shown in Fig.~\ref{fig:9}(b). The corresponding vertical displacement is then given by
\begin{align}\label{eq:mapping_y}
\frac{\Delta y}{y}=&\frac{y_{eff}-y}{y}=\frac{l_0\tan(\alpha^{case 3}_{eff})-l_0\tan(\alpha^{case 3})}{l_0\tan(\alpha^{case 3})}\nonumber\\
=&\frac{\tan(\alpha^{case 3}_{eff})-\tan(\alpha^{case 3})}{\tan(\alpha^{case 3})}.
\end{align}
These results are shown in Fig.~\hyperref[fig:9]{\ref*{fig:9}(c)}, where it is clear that substantial vertical shifts are needed to account for the suppression of $J_{12}$ in the triangular triple dot case. For example, when $\alpha=\ang{30}$, $\Delta y/y\approx0.95$, corresponding to a $\sim95$\% vertical shift. From Fig.~\ref{fig:5}, where $y=l_0\tan\alpha=16.17$ nm, we see that $y_{eff}\approx25$ nm, yielding $\Delta y/y\approx 0.55$. Although this is less than 0.95, it is still large enough that we believe this is the primary mechanism responsible for the suppression of $J_{12}$ in the triangular case. The discrepancy is likely due to the downward shift of the electrons in D1 and D2, which we have neglected in this analysis. From Fig.~\ref{fig:5}, we see that this shift is on the order of 5 nm; this would bring the net vertical shift up to $y_{eff}\approx 30$ nm, which is consistent with the 95\% value obtained from our analysis. As the angle or $l_1$ gets larger, the Coulomb repulsion becomes weaker, and the vertical displacement becomes negligible, which is also clear from Fig.~\ref{fig:9}(c). One can also see this from the shifts in the single-particle densities at larger $\alpha$ shown in Appendix~\ref{app:A}.

Before we conclude, it is worth commenting on why we have not seen any evidence of negative exchange interactions, even though our previous work showed that these can arise in quantum dot systems containing as few as four electrons \cite{Deng2018}. In that work, we showed that if four electrons are confined in a symmetric parabolic potential, the ground state is a triplet provided the splitting between the second and third single-particle levels is sufficiently small. This splitting vanishes in the limit of zero magnetic field due to rotational symmetry, and it remains small in the low magnetic field regime. These findings suggest that, in the present case of the linear triple dot, if one were to gradually tune $l_1$ down to zero, one would see the exchange energy reach a maximimum positive value and then decrease all the way down to negative values as the three dots merge into one big dot. Before $l_1$ reaches zero, the triple dot potential looks like one large elliptical dot. In Ref.~\cite{Deng2018}, we calculated how the exchange energy depends on dot ellipticity, and we showed that a transition from negative to positive exchange occurs as the ellipticity increases past a certain threshold value that depends on the confinement energy. This transition happens because increasing the ellipticity breaks the rotational symmetry of the dot and opens a gap between the second and third single-particle levels. Fig. 3 of Ref.~\cite{Deng2018} shows that the exchange energy vanishes when the ellipticity $\hbar\delta\omega$ is about 1-2 meV (when the confinement energy is 7-8 meV), which corresponds to a difference of 2-4 nm between the vertical and horizontal extent of the dot potential. For the linear triple dot geometry, this implies that the small dots would need to overlap almost completely with the mediator, which happens for $l_1\lesssim 5$ nm, which is well below the $l_1$ values we have considered. Thus, it is not surprising that we have not encountered negative exchange energies in this work. Unfortunately, probing this crossover behavior from three separate dots to one large dot is computationally challenging because a very large number of single-particle basis states would be needed to obtain accurate results, which translates to a very large computational cost. It would also be interesting to explore the possibility of negative superexchange interactions when four electrons are confined to the mediator instead of two. We leave these investigations to future work.

\section{Conclusions}\label{sec:conclusions}
In this work, we explored the interplay of normal exchange and superexchange processes in triple quantum dot systems where a large dot is used to mediate long-range spin-spin interactions between a pair of smaller dots. We consider triangular geometries in which both normal exchange and superexchange can be present simultaneously. Using configuration interaction simulations, we showed that the effective exchange energy receives a modest enhancement due to superexchange when the mediating dot is brought sufficiently close to the small dots. We further showed that this enhancement can be increased by two orders of magnitude if the mediating dot is loaded with two electrons, a phenomenon we attribute to a combination of Fermi statistics and quantum confinement. We also found that the effective exchange energy exhibits non-monotonic behavior as the distance between the small dots and the mediator is varied. This can be understood as a consequence of the rapid decay of superexchange with distance and a more slowly changing lateral shift of the electron positions in the small dots due to Coulomb repulsion from the mediator electrons. Our calculations also reveal that the effective exchange interaction can be made still larger by placing the mediating dot exactly between the two smaller dots, and we provided evidence that the somewhat smaller interaction in the triangular case is likely due to an additional Coulomb repulsion that is not present in the linear case. Moreover, we found that in addition to sharply increasing the effective exchange coupling, the electron-filled mediator also more than doubles the range of the interaction.

Our results show that including electrons in a quantum dot mediator can substantially enhance the strength and range of spin-spin interactions between remote quantum dots. They also suggest that the precise geometry of the dots can have important ramifications and provide additional flexibility in the design of larger-scale architectures based on quantum dot-mediated exchange couplings. 

\section*{Acknowledgments}
This work is supported by the Army Research Office (W911NF-17-0287) and by the U.S. Office of Naval Research (N00014-17-1-2971).

\bibliographystyle{apsrev4-1}

\bibliography{library}

\appendix

\section{Single-particle densities and mediator detuning values}\label{app:A}
In this appendix, we show additional plots of the single-particle density for cases 2, 3 and 4, for both the lowest-energy singlet-like (S) and triplet-like (T) eigenstates with total spin projection $S_z^{total}=0$. We also provide further details about the way we choose our potential cuts and detuning values $\Delta$ to guarantee the mediator contains the desired number of electrons in each case.

\subsection{Triangular triple dot with two electrons (case 2)}
In this case, the potential cut between D1 and D2 is always the half-line, $x=0$, $y<y_0$, and the horizontal line, $y=y_0$. For the latter, we choose different values of $y_0$ depending on the angle we choose. For $\alpha=\ang{30},\ang{40},\ang{45},\ang{50}$, we set $y_0=6, 10, 12, 15$ nm, respectively. For the larger angles, we do the following. First we choose the detuning, $\Delta$, and then determine the equal-potential point on the line connecting D1 and D3. Due to symmetry, we can find a similar point on the line connecting D2 and D3. We then connect these two points by a line, which gives us the $y=y_0$ cut. The reason we do not use this procedure for smaller angles is because the big dot is very close to the $x$-axis, and the equal-potential point is too close to the small dots in these cases, and so placing the cut here would leave the small dots with too much overlap in the big dot region.

\begin{figure*}[!tbp]
\centering
{{\includegraphics[trim=0cm 0cm 0cm 0cm, clip=true,width=17cm, angle=0]{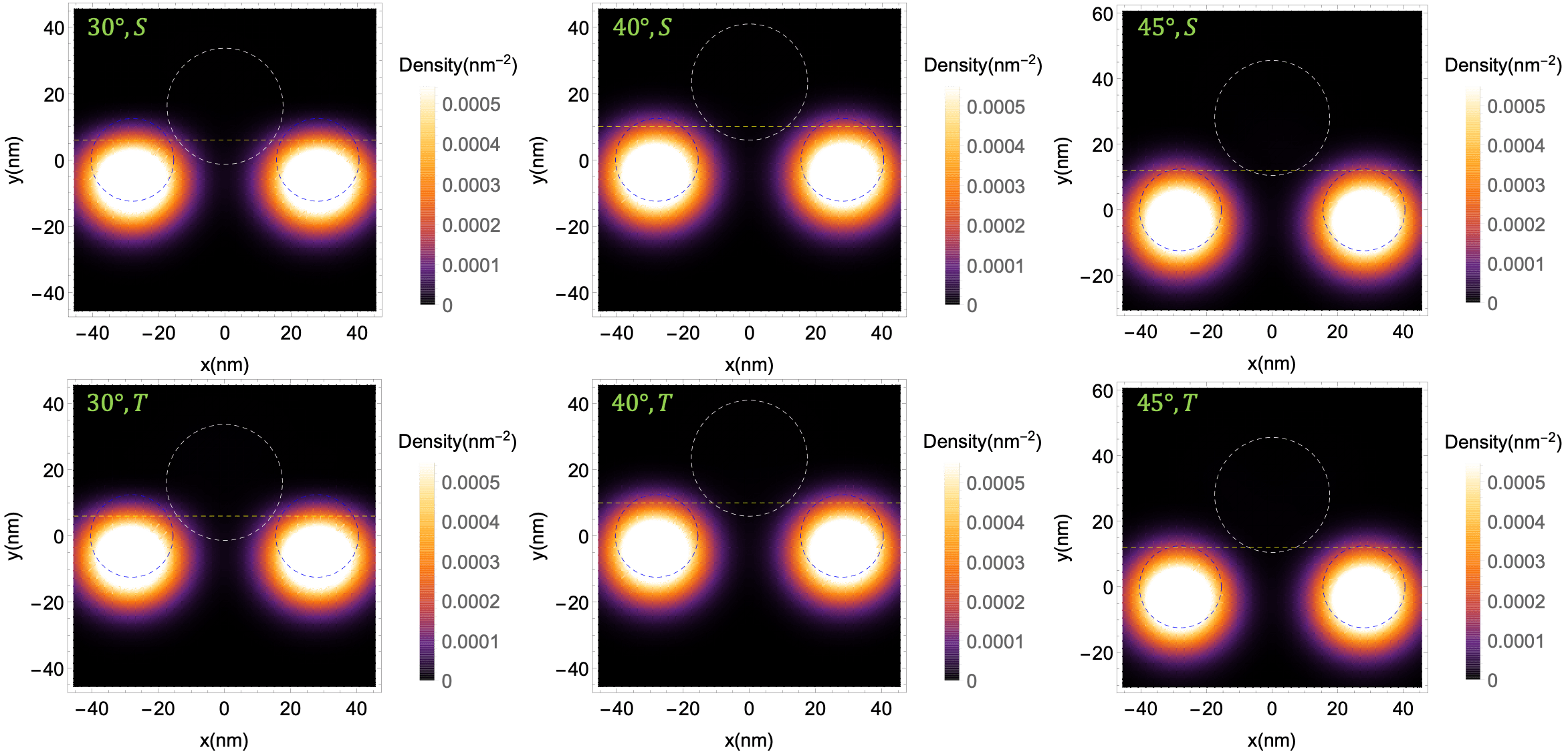}}}\\
{{\includegraphics[trim=0cm 0cm 0cm 0cm, clip=true,width=17cm, angle=0]{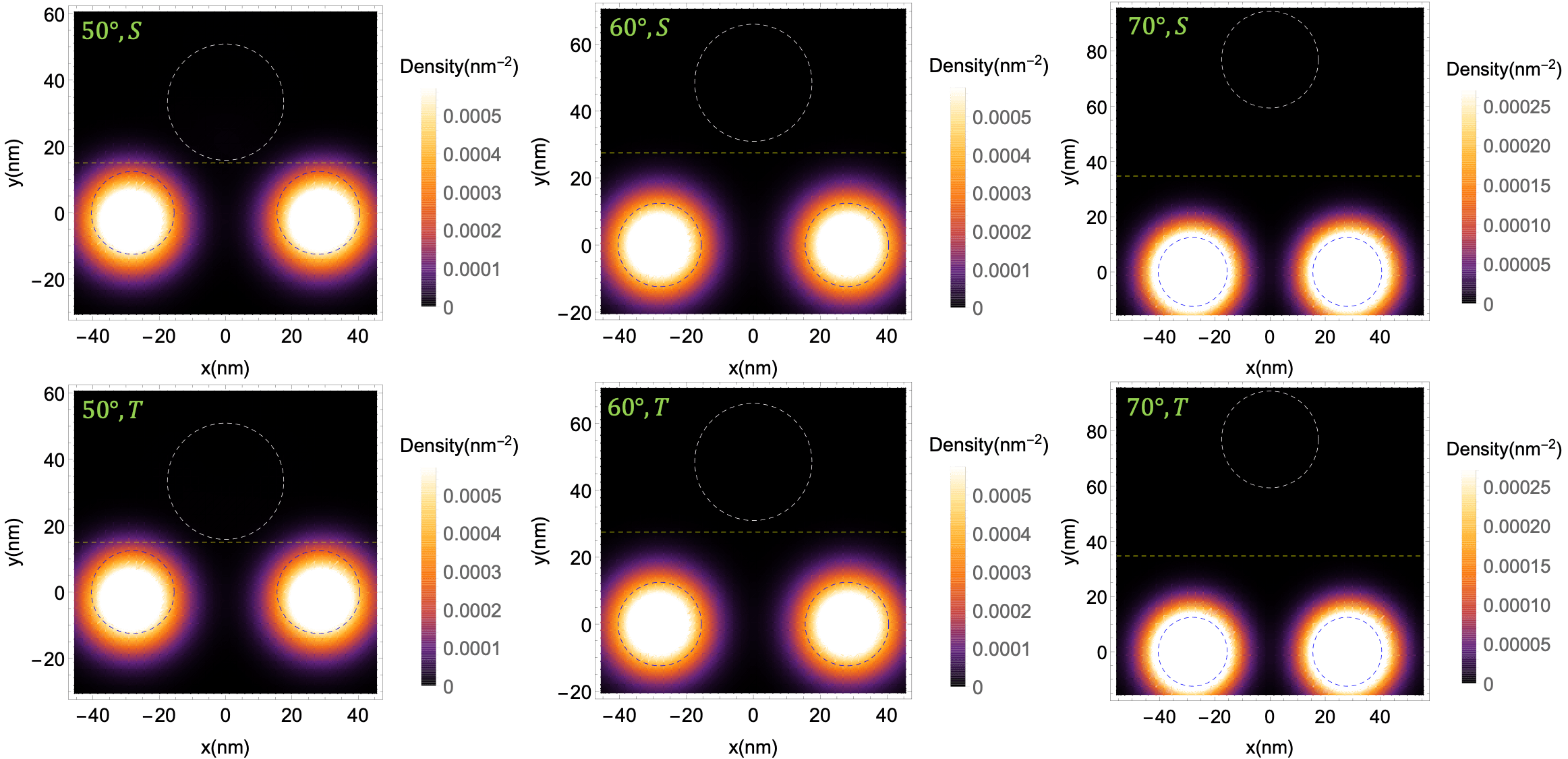}}}
\caption{Single-particle density plots for the triangular triple dot with two electrons (case 2) for $\alpha=\ang{30},\ang{40},\ang{45},\ang{50},\ang{60},\ang{70}$ and for the lowest-energy singlet-like (S) and triplet-like (T) states. The white and blue dashed circles indicate the dot potentials, and the yellow dashed line is the potential cut (here we do not show the $x=0$ cut in the plots). The big dot contains approximately zero electrons. At low angles, the two small dots move a little bit downward due to the large detuning of the big dot. At high angles, this effect is small because the dots are sufficiently far apart.}\label{app_fig.1}
\end{figure*}

We set the detuning on D3 to a high value, $\Delta=20$ meV, to deplete the mediator as much as possible. The tolerance threshold for the electron number on D3 is set as discussed in Sec.~\ref{sec:Triangular_two_electrons}. The corresponding single-particle density plots for several different angles are shown in Fig.~\hyperref[app_fig.1]{\ref*{app_fig.1}}. The three dot potentials are marked by white and blue dashed circles, and the potential cut $y=y_0$ is indicated by a yellow dashed line. It is evident that the mediator remains empty in all cases.

\subsection{Triangular triple dot with four electrons (case 3)}
Here, we do the same potential cuts as in case 2. To determine appropriate choices for the detuning $\Delta$, we perform a systematic scan over $\Delta$ values, in each case calculating how many electrons are in the big dot region above $y=y_0$. We find that the following values correspond to having two electrons in the big dot: $\Delta=1.0$, $-1.0$, $-1.5$, $-2.0$, $3.5$, $4.0$, $4.0$, $4.0$, $4.0$, $4.0$, $4.0$, $4.0$ meV for $\alpha=\ang{30}$, $\ang{40}$, $\ang{45}$, $\ang{50}$, $\ang{60}$, $\ang{64}$, $\ang{68}$, $\ang{70}$, $\ang{72}$, $\ang{74}$, $\ang{76}$, $\ang{80}$, respectively.

\begin{figure*}[!tbp]
\centering
{{\includegraphics[trim=0cm 0cm 0cm 0cm, clip=true,width=17cm, angle=0]{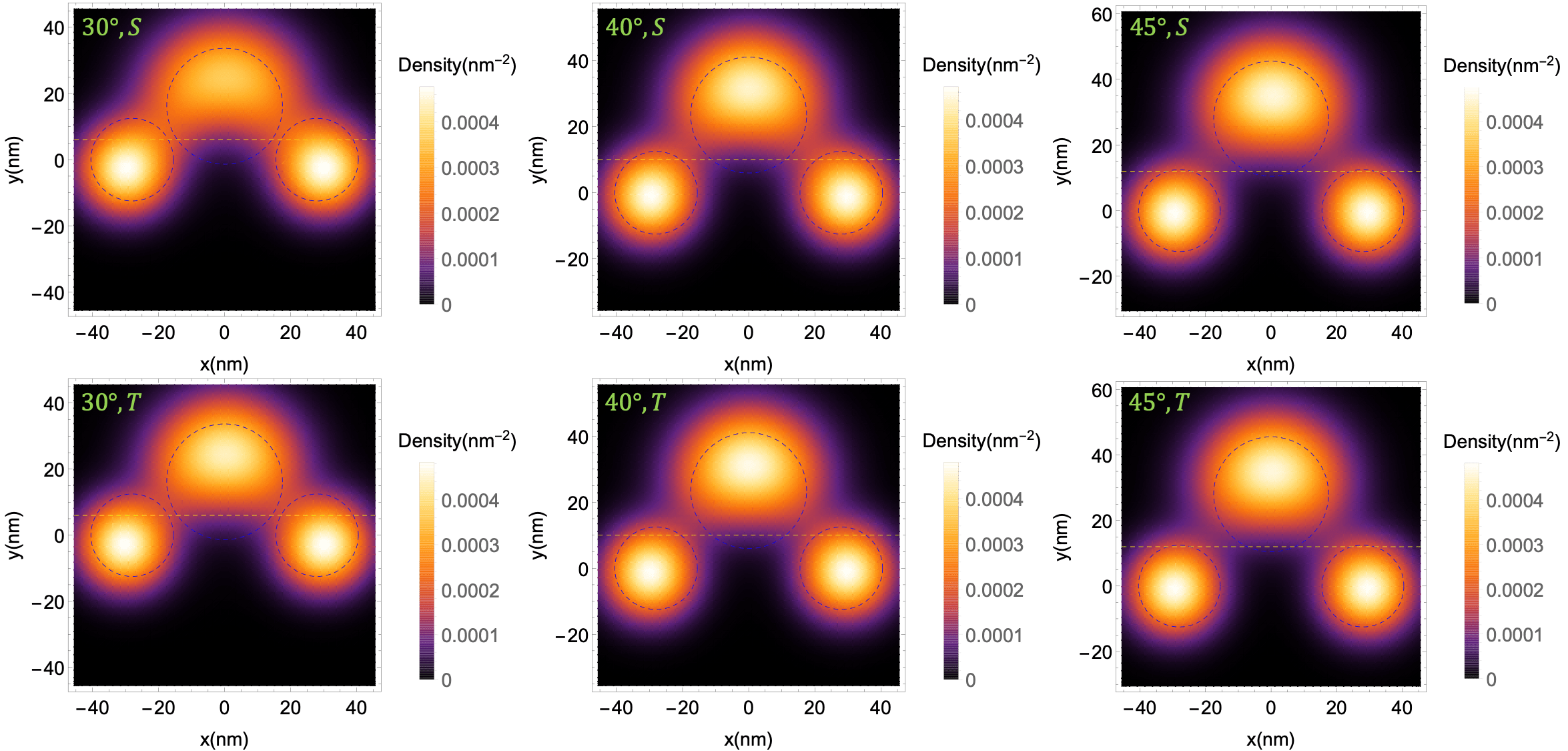}}}\\
{{\includegraphics[trim=0cm 0cm 0cm 0cm, clip=true,width=17cm,angle=0]{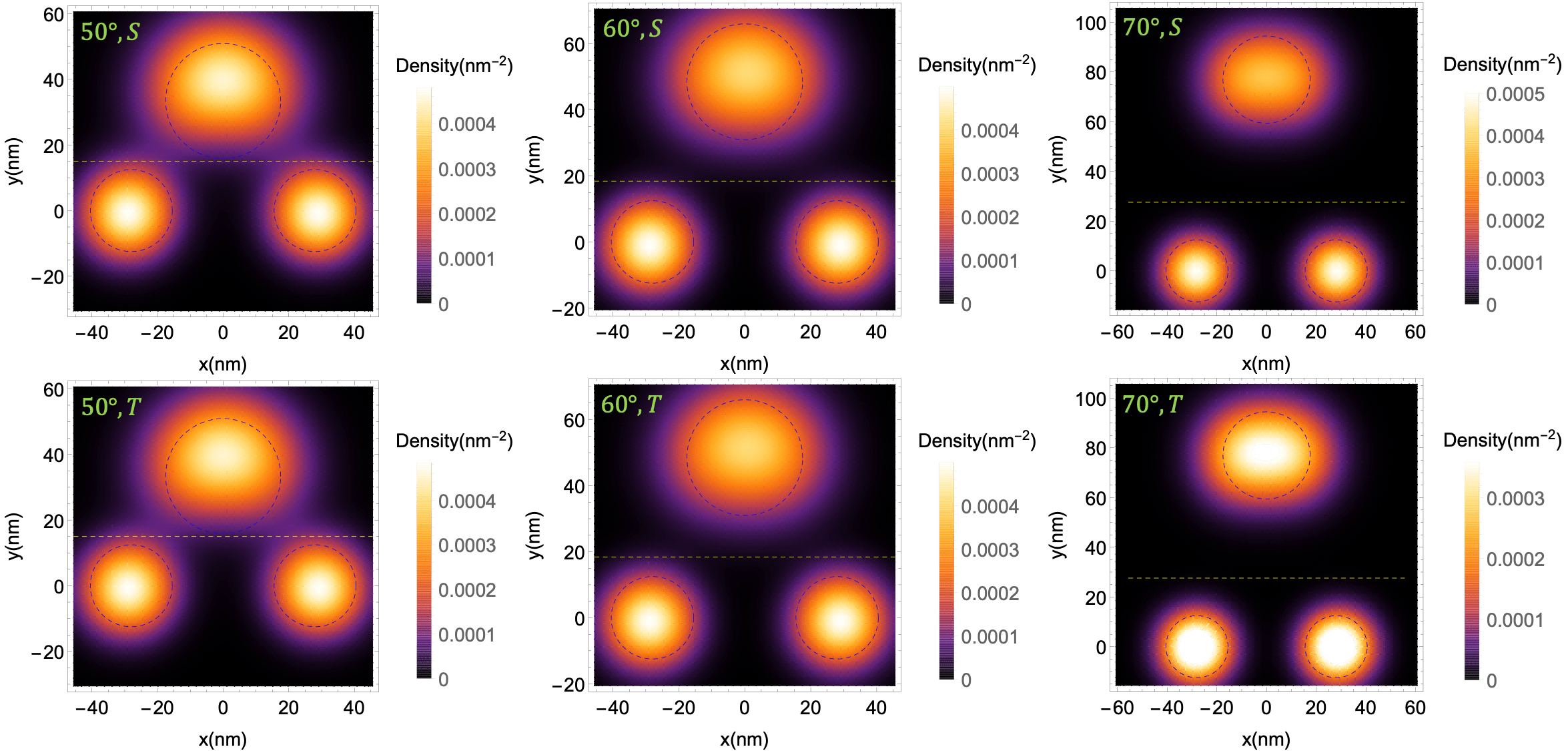}}}
\caption{Single-particle density plots for the triangular triple dot with four electrons (case 3) for $\alpha=\ang{30},\ang{40},\ang{45},\ang{50},\ang{60},\ang{70}$ and for the lowest-energy singlet-like (S) and triplet-like (T) states. The white and blue dashed circles indicate the dot potentials, and the yellow dashed line is the potential cut (here we do not show the $x=0$ cut in the plots). In each case, the big dot contains two electrons as can be confirmed by integrating the density over the upper region. The position shifts of the electrons in the big dot are significantly larger for smaller angles compared to larger angles, which confirms the results obtained in Sec.~\ref{sec:Linear_four_electrons} and Fig.~\hyperref[fig:9]{\ref*{fig:9}}.}\label{app_fig.2}
\end{figure*}

We show several density plots for various values of $\alpha$ in Fig.~\hyperref[app_fig.2]{\ref*{app_fig.2}}. The three dots are illustrated by the blue dashed circles, and the potential cut at $y=y_0$ is indicated by the yellow dashed line. One can see that the shifts in the positions of the electrons in the big dot are much larger at lower angles, which agrees with the analysis in Sec.~\ref{sec:Linear_four_electrons} and  Fig.~\hyperref[fig:9]{\ref*{fig:9}}.

\subsection{Linear triple dot with four electrons (case 4)}
In this case, we make two potential cuts $(x=\pm x_0)$ parallel to the $y$-axis. To do this, we first choose the detuning $\Delta$ for D3, and then compute the equal-potential point $(x_0,0)$ between D2 and D3. Thus, we can separate the space into three parts using the two cuts $x=\pm x_0$.

To decide the detunings $\Delta$, we again scan over a range of values and integrate the density to see how many electrons are in the big dot region (middle region). This process yields the following values at which two electrons are confined to D3: $\Delta=-0.5$, $0.0$, $0.0$, $0.0$, $0.0$, $0.0$, $2.0$, $3.0$, $3.5$, $3.5$, $3.5$, $3.5$ meV for $\alpha=\ang{30}$, $\ang{35}$, $\ang{40}$, $\ang{45}$, $\ang{50}$, $\ang{55}$, $\ang{60}$, $\ang{64}$, $\ang{68}$, $\ang{70}$, $\ang{72}$, $\ang{80}$, respectively.

\begin{figure*}[!tbp]
\centering
{{\includegraphics[trim=0cm 0cm 0cm 0cm, clip=true,width=17cm, angle=0]{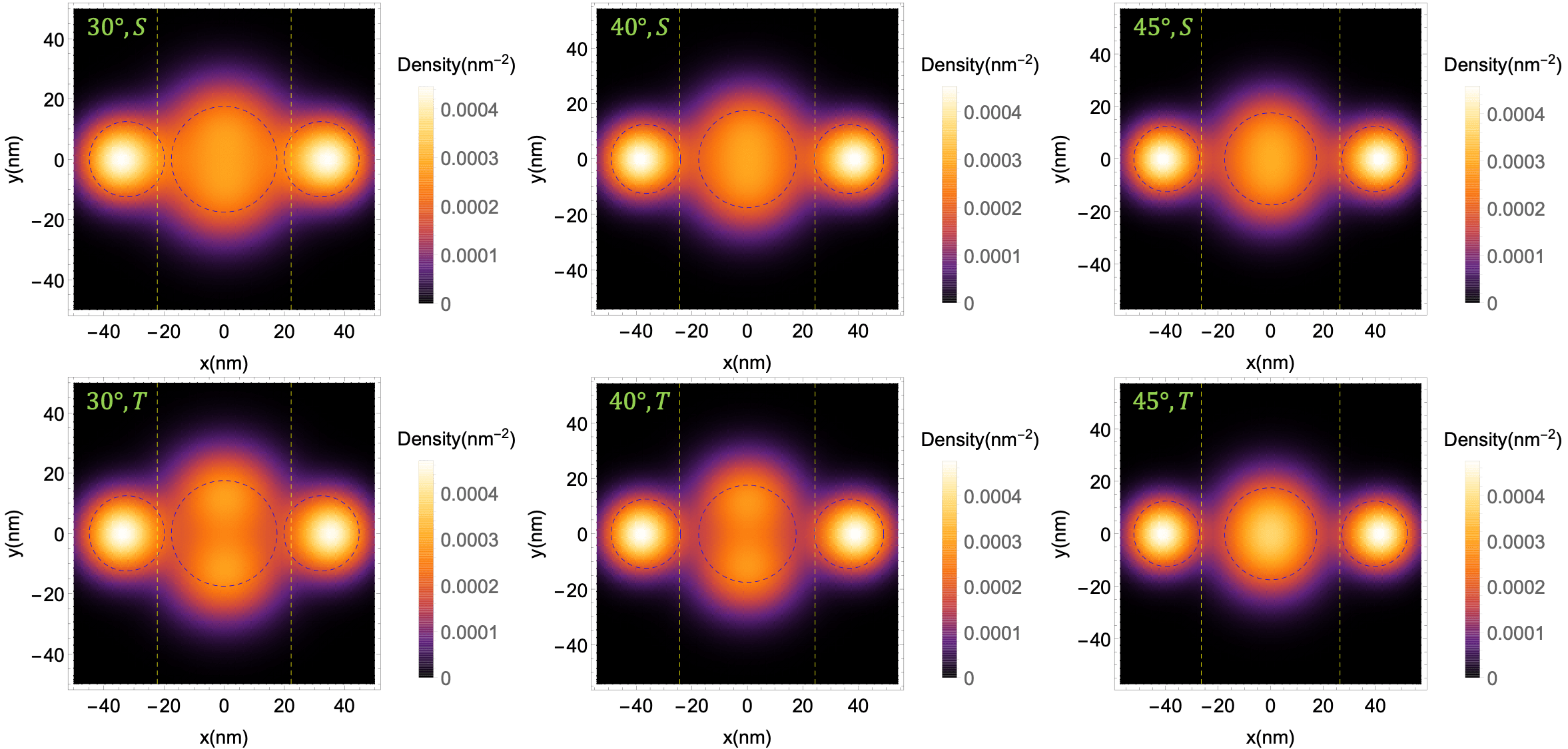}}}\\
{{\includegraphics[trim=0cm 0cm 0cm 0cm, clip=true,width=17cm, angle=0]{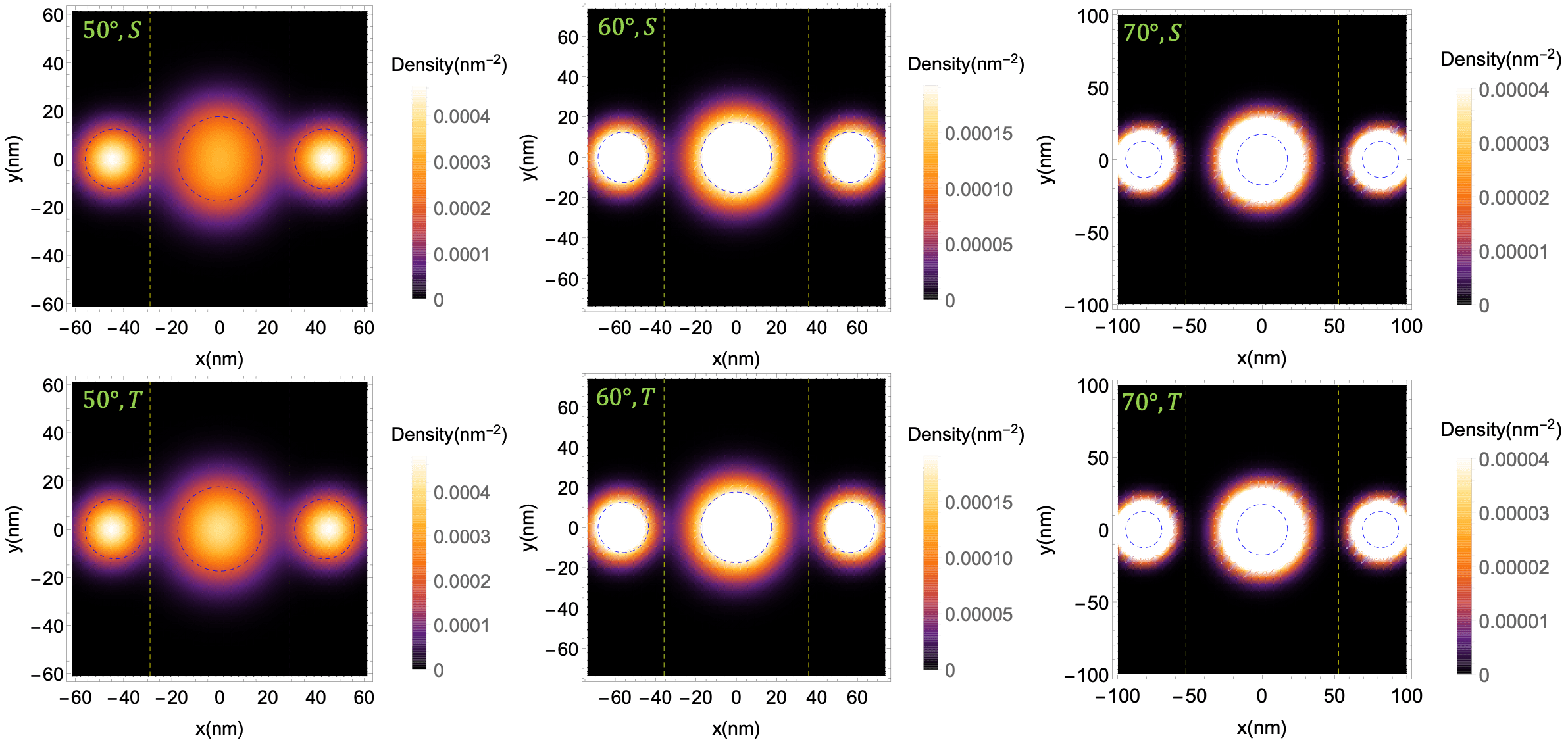}}}
\caption{Single-particle density plots for the linear triple dot with four electrons (case 4) for $\alpha=\ang{30},\ang{40},\ang{45},\ang{50},\ang{60},\ang{70}$ and for the lowest-energy singlet-like (S) and triplet-like (T) states. The blue dashed circles indicate the dot potentials, and the yellow dashed lines are the potential cuts. The big dot contains two electrons as can be confirmed by integrating the density over the middle region.}\label{app_fig.3}
\end{figure*}

Single-particle density plots for several different values of $\alpha$ are shown in Fig.~\hyperref[app_fig.3]{\ref*{app_fig.3}}. The three dots are illustrated by the blue dashed circles, and the potential cuts at $x=\pm x_0$ are indicated by the yellow dashed lines.

\end{document}